# Silk-templated Nanostrips as Superprotonic Fibre Sensors

Jianhui Zhang[1,2], Ahmed Salem[1,2,3], Robert Tidswell[4], Haowei Wang[5,6], Vikramjeet Singh[7], Laurence B. Lovat[2,4], Manish K. Tiwari[1,2,7]*

[1]Nanoengineered Systems Laboratory, UCL Mechanical Engineering, University College London, London, WC1E 7JE, UK.
[2]Hawkes Institute, University College London, London, W1W 7TS, UK.
[3]UCL Medical Physics and Biomedical Engineering, University College London, London, WC1E 7JE, UK.
[4]Department of Critical Care, University College London Hospitals NHS, NW1 2BU, UK.
[5]Centre for Precision Healthcare, UCL Division of Medicine, University College London, London, WC1E 6JF, UK.
[6]Division of Biomaterials and Tissue Engineering, Royal Free Hospital, University College London, London, NW3 2PF, UK.
[7]Manufacturing Futures Lab, Mechanical Engineering, University College London, London, E20 2AE, UK.

***Corresponding author. Email: m.tiwari@ucl.ac.uk**

## Abstract

Transforming textile fibres into sensors can facilitate continuous physiological monitoring for improving human healthcare. Coating fibres with electronic conductors can help detect physiological stimuli but their sensitivity scales with thickness and often lowers the fibre mechanical flexibility. Proton conductors are potential alternatives; however, they suffer from fragile physical interfaces, and sluggish kinetics. Here, we report a bio-templated, scalable strategy to create superprotonic interfaces using natural silk fibres. We exploit ordered nanofibrils on silk fibres to selectively grow continuous metal-organic framework (MOF) nanostrips. This patterned, ca. 10 μm-diameter fibre-structure provides low-defect interfacial pathways for ultrafast proton transport, yielding conductivities up to 40 S $cm^{-1}$ – two orders of magnitude above the state of the art – with millisecond response and high mechanical robustness. The individual micro-thread sensors facilitate respiratory monitoring with a mean bias of 0.01 breaths $min^{-1}$ in volunteer testing, unaffected by human movement and readily capture breathing signature in high-flow oxygen therapy devices. We integrate arrays of micro-thread sensors in textiles and exploit deep learning to demonstrate high-resolution spatiotemporal humidity mapping in dynamic flow-fields and resolve asymmetric respiratory patterns that elude conventional single-point detectors. Our results should have broad-ranging implications, e.g. in wearable health monitoring, bioelectronics, and human-machine interfaces.

* Corresponding author, email: m.tiwari@ucl.ac.uk, phone: +44 20 3108 1056

## Main

Superprotonic conductors are a class of materials that achieve proton conductivities, typically $10^{-4}$-$10^{-2}$ S $cm^{-1}$, through continuous hydrogen-bonded networks of adsorbed water or acidic groups[1-3]. These conductors have shown remarkable potential in wearable sensing, fuel cells and electrolysis[4-7], where charge is carried by protons. Within the bulk and at the interface, these protons move along the networks by structural hopping between adjacent sites or by diffusing as hydrated carriers[1]. In either case, breaks or defects in the network interrupt conduction. Since most superprotonic conductors are processed as defect-rich powders or disordered polymer films[8-10], their proton conductivity is limited and their use as wearables presents several challenges. The amorphous structure and hygroscopic swelling of soft proton-conducting polymers make conduction slow and unstable[11,12]. Rigid crystalline materials such as metal-organic frameworks (MOFs) provide higher intrinsic conductivity[2,13]. However, they are usually deposited as powdery or thick films [8,10], in which grain boundaries and discontinuities suppress proton transport (Extended Data Fig. 1). Due to the resulting poor conductivity and sensitivity, an enclosed, moisture-enriched environment (i.e. a mask) is often required [14,15], which is uncomfortable to wear and impractical for continuous monitoring [16]. Moreover, such coatings with low strain tolerance delaminate and crack on flexible substrates under repeated deformation [10,17].

To address these issues, we aimed to undertake a templated growth strategy for well-controlled growth of superprotonic conducting interface. Nature uses proteins with patterned hydrophilic–hydrophobic surfaces to template the oriented growth of inorganic crystals, called biomineralisation [18,19]. In fact, silk fibres with naturally occurring multi-scale hierarchical structures [20-22] could serve as an ideal bio-template to guide epitaxial MOF growth, using standard techniques[23] (Fig. 1 a, b). This templating produces continuous MOF nanostrips aligned along the fibre, creating highly conductive proton-transport pathways while preserving the flexibility of the underlying fibres (Fig. 1c–e). The MOF nanostrips reach a superprotonic conductivity about two orders of magnitude higher than previously reported (up to ~40 S $cm^{-1}$) [1,24], together with a millisecond response and high sensitivity. These properties make a micro-thread yarn sufficient for imperceptible breathing monitoring (Fig. 1 f) and scalable sensor arrays (Fig. 1 g) capable of spatiotemporal imaging of humidity fields.

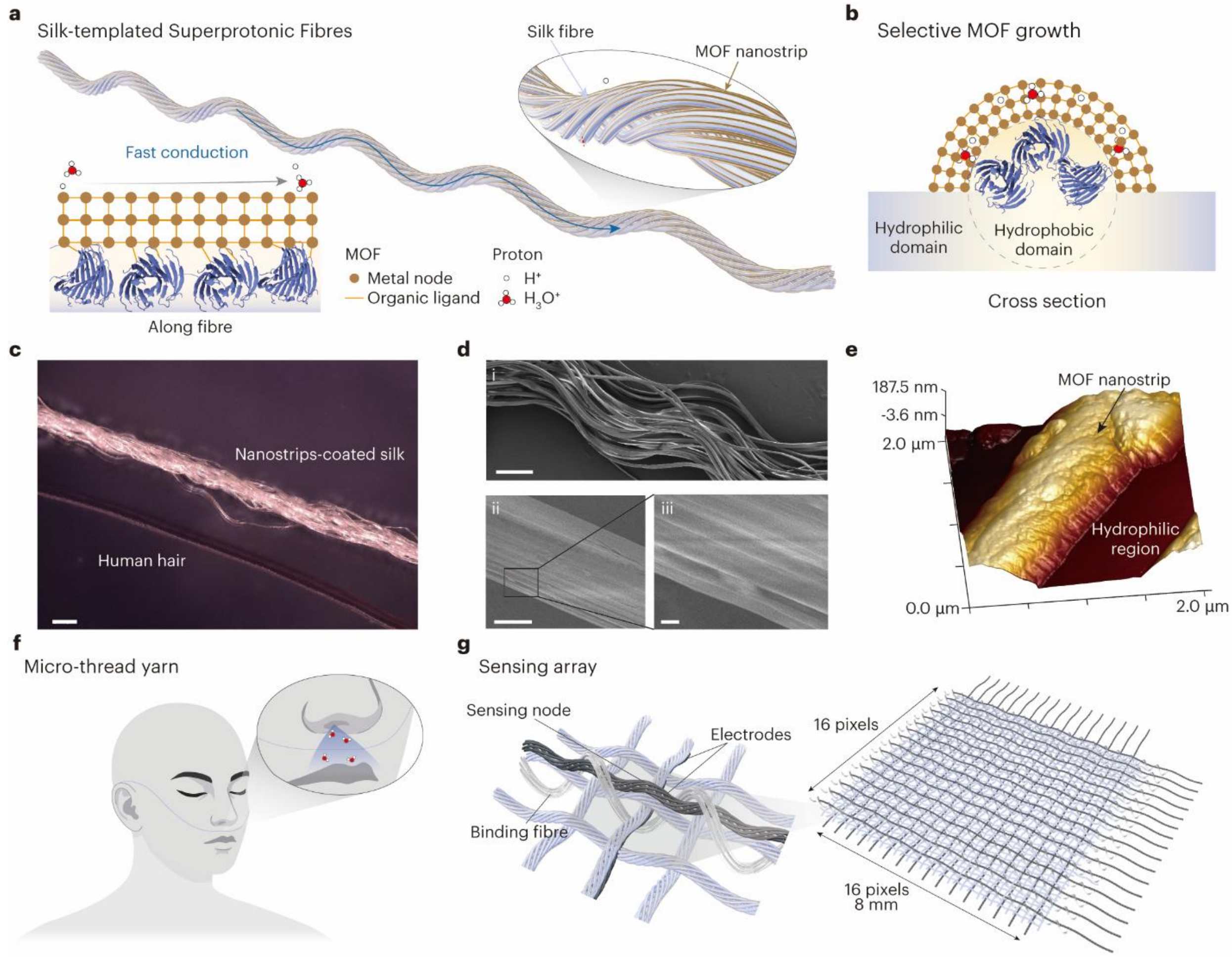

**Figure 1. Silk-templated MOF nanostrips for superprotonic humidity sensing.**
(a) Schematic of superprotonic conductive fibres based on silk fibres coated with MOF nanostrips. Inset (left) shows MOF nanostrips aligned along the fibre, enabling fast proton conduction. (b) Cross-sectional schematic illustrating selective growth of thin MOF nanostrips on hydrophobic , β-sheet-rich domains, producing a heterogeneous nanostructured coating directed by the multiscale silk template. (c) Optical microscopy image showing a MOF nanostrip-coated silk yarn next to a human nasal hair. Scale bar: 200 µm. (d) Representative Scanning electron microscopy (SEM) images of MOF nanostrip-coated silk fibres from independently prepared samples show axially aligned nanostrip textures along the fibre surface. Scale bars: (i) 100 µm, (ii) 4 µm and (iii) 500 nm. (e) Atomic force microscopy (AFM) topography image of the MOF nanostrip-coated silk surface, revealing crystalline nanostrip domains separated by smoother protein-rich hydrophilic regions. PeakForce nanomechanical maps (see Supplementary Fig. 3) provide further contrast between these domains. (f) Schematic of breathing monitoring using a micro-thread yarn worn under the nose. (g) Schematics of a high-density textile humidity-sensing array assembled by weaving or embroidering conductive fabric electrodes across the superprotonic conductive fabric. This strategy enables compact arrays, including a 16 × 16 pixel array within an 8 × 8 mm area.

## Design and Characterisation

Degummed silk fibres are ~10 μm in diameter, which is finer than a human hair (Fig. 1c). Each fibre consists of fibroin filaments organised into aligned nanofibrils, in which β-sheet-rich crystalline domains alternate with more amorphous hydrophilic regions [20]. We envisioned that this multiscale hierarchy, together with the chemical heterogeneity of silk, could serve as a natural template to spatially direct MOF growth on a soft fibrous substrate.

To test this idea, silk textiles were first subjected to a simple organic-solvent treatment that enriched the β-sheet structure [21] and exposed the nanofibrillar surface (Supplementary Fig. 2). MOF linkers were then covalently anchored to the protein backbone through an EDC/NHS-mediated reaction, followed by layer-by-layer (LbL) growth of UiO-66-$NH_2$, a zirconium-based MOF chosen for its moisture stability and hydrophilic ligands [13]. Two control coatings, MOF nanofilm and MOF nanopowder, were prepared by solvothermal growth and nanoparticle spraying, respectively (Extended Data Fig. 1). After three assembly cycles, the bio-templated growth process preserved the structural integrity of the silk substrate while generating aligned nanostrip coatings along the fibre surface (Fig. 1d), in sharp contrast to the rougher control coatings (Extended Data Figs. 1). Atomic force microscopy (AFM) further confirmed this nanostrip morphology (Fig. 1e).

Zooming into the nanostrip regions, the AFM further resolved a smooth, nanoscale morphology (Fig. 2a). Nanomechanical mapping also provided evidence of heterogeneous surface organisation, distinguishing MOF-coated domains from intervening protein-rich regions, and a nanostrip structure of MOF on fibres (Supplementary Note 2 and Supplementary Fig. 3). As the number of assembly cycles increased, these aligned features gradually disappeared and the surface evolved towards a thicker, more disordered morphology resembling the nanofilm control (Extended Data Fig. 2) - the nanostrip architecture is achieved within a narrow assembly window.

To determine the dimensions of the nanostrips, the silk substrate was dissolved to release the deposited MOF structures, which were then deposited on mica for AFM analysis. The resulting nanostrips were ultrathin, with a thickness of ~70–80 nm (Fig. 2b and 2c), comparable to the lower end of previously reported MOF nanofilms [23] but here formed directly on a flexible fibre template. Despite the hydrophilic chemistry of UiO-66-$NH_2$, water droplets on the nanostrip-coated textile remained in a non-wetting state, with an advancing contact angle of ~146° (Fig. 2d), indicating that the aligned multiscale topography strongly altered the apparent wetting behaviour of the fibre surface [29] (see quantitative analysis in Supplementary Note 1). Thermogravimetric analysis (TGA) showed a low final residue of ~1.9 wt% (Fig. 2e), corresponding to an estimated initial MOF loading of ~6.7 wt% and confirming that the coating was conformal and lightweight rather than a thick particulate overlayer.

Structural characterisation verified the successful formation of UiO-66-$NH_2$ on silk without destroying the underlying fibroin framework. X-ray diffraction (XRD) showed the coexistence of characteristic silk II and UiO-66-$NH_2$ reflections [30,31] (Fig. 2f), while Fourier-transform infrared (FTIR) spectroscopy and elemental mapping confirmed the presence of the MOF coating on the fibre surface [32] (Fig. 2g and Supplementary Fig. 4). Solvent treatment and linker exposure also increased the β-sheet character of silk (Fig. 2h and Supplementary Fig. 5), consistent with the formation of a more ordered and exposed fibrous bio-template before MOF assembly. The pretreated and coated yarns exhibited sharp brittle fracture rather than the more ductile failure of pristine silk (Fig. 2i and Supplementary Fig. 6), indicating reduced plastic deformation and a shift towards a more rigid, ordered failure mode [17]. This change is consistent with the increased β-sheet character induced by pretreatment [21], together with the presence of a bonded MOF nanostrip coating on the fibre surface. Together, these results show that silk-templated assembly produces an ultrathin, continuous and structurally ordered MOF nanostrip that is distinct from conventional powder- or film-based coatings [1].

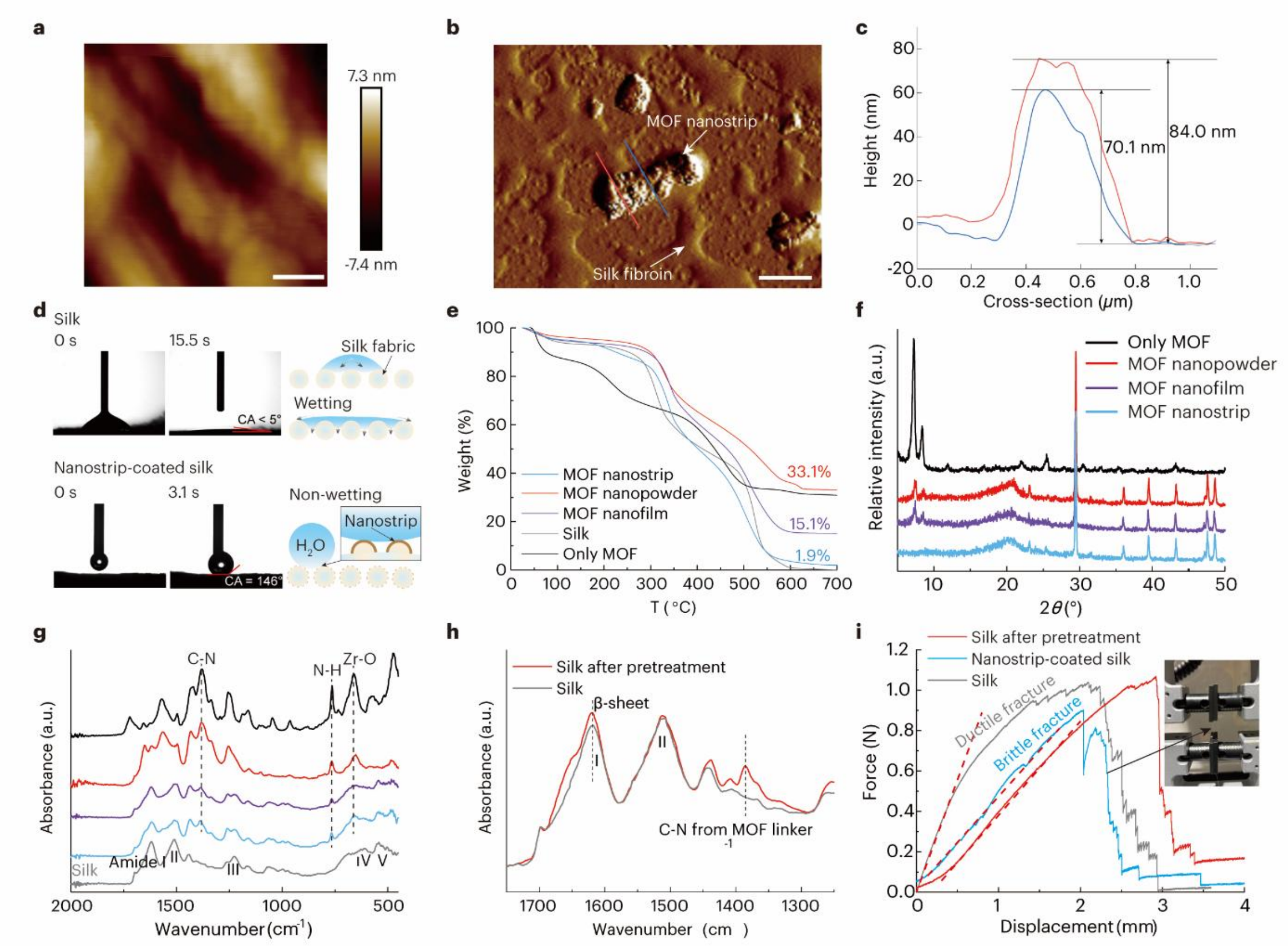

**Figure 2. Characterisation of MOF nanostrips**.

(a) Surface morphologies of MOF nanostrip, fabricatged via LbL assembly (3 cycles). Scale bar: 100 nm. (b) AFM deflection map of an individual segment of a MOF nanostrip and (c) its corresponding cross-sectional height profile. These samples were obtained by dissolving the silk substrate and depositing the liberated nanostrips onto a flat mica substrate. Scale bar, 500 nm. (d) Advancing contact angle images of ~10 μL water droplets on pristine silk textile and MOF nanostrip-coated silk textile, with corresponding wetting schematics. The coated textile exhibits a Cassie-like non-wetting state. (e) TGA curves of different MOF-coated silk samples, showing water mass loss at the first stage and low high-temperature residue for the nanostrip-coated sample, consistent with a conformal but ultrathin MOF coating. (f) XRD patterns of MOF nanoparticles and different MOF-coated silk fibres, confirming the crystalline silk and MOF phases. Diffraction peaks near 7.3° and 8.3° are assigned to the (111) and (002) planes of UiO-66-$NH_2$, whereas the broad reflection at ~20° corresponds to silk II. (g) FTIR spectra of MOF nanoparticles and MOF-coated silk fibres, confirming successful formation of the MOF coating. The FTIR peaks at 1384, 768 and 663 $cm^{-1}$ confirm the presence of aromatic amine and Zr–O vibrations associated with UiO-66-$NH_2$. (h) FTIR spectra of silk after pre-treatment with the linker solvent, showing the appearance of linker-related C–N bands. The enhanced amide I band near 1620 $cm^{-1}$ after solvent treatment indicates increased β-sheet character. (i) Tensile test on single yarns, showing distinct fracture modes before and after coating. Fracture behaviour was examined across at least three single yarns per group as shown in Supplementary Fig. 6. Slope of dashed lines (fitting early part of the load deformation curves) indicate modulus.

## Interfacial superprotonic conduction

To investigate whether our bio-templated nanostrip structure improved proton transport and humidity sensing, we compared UiO-66-$NH_2$ prepared by three routes: templated nanostrip on silk, solvothermally grown nanofilm and sprayed nanopowder coatings. Electrochemical impedance spectra recorded under controlled humidity showed that, as relative humidity increased from ambient conditions to ~83%, all samples evolved from highly resistive behaviour towards proton-conducting responses, but the nanostrip-coated fibres consistently exhibited much lower impedance than both controls (Fig. 3a and Extended Data Fig. 4). The nanostrip samples also operated over the broadest humidity window (Fig. 3b), compared with the nanofilm and nanopowder controls. Equivalent-circuit fitting further showed that the charge-transfer resistance and the inductance of the nanostrip samples decreased steeply with humidity (Supplementary Note 4 and Supplementary Fig. 10). Together, these results indicate that the templated nanostrip morphology enables efficient proton transport across the humidity range relevant to human breathing.

We then examined the dynamic response of the nanostrip architecture under rapid humidity modulation. Under oscillatory humidity cycling at ~5 Hz, the nanostrip fibres showed response and recovery times of ~34 ms and ~78 ms, respectively (Fig. 3c), and could still track humidity fluctuations at up to 50 Hz, albeit with some signal drift (Fig. 3d). These response rates are comparable to the fastest reported humidity-sensitive low-dimensional materials (e.g. graphene oxide [33]), but are achieved here in a covalently bonded thin coating on fibre rather than a fragile thick coating, with better durability. The superior performance was also maximised with low thickness at only 3 assembly cycles, consistent with the structural evolution observed in Extended Data Fig. 4b.

This behaviour could be further tuned through MOF chemistry. We systematically varied the linker polarity (pristine, –$NH_2$, –OH) and the metal node (Zr-based UiO-66, Zn-based MOF-5) (Extended Data Fig. 4). Among them, UiO-66-OH delivered the highest sensitivity, the lowest resistance and the highest proton conductivity, reaching ~40 S $cm^{-1}$ under humid conditions (calculated as detailed in Supplementary Note 5). This is attributed to the symmetric –OH groups on both sides of the linker, which provide a denser interfacial hydrogen-bond network, as supported by our molecular dynamics simulations (Extended Data Fig. 5b). By contrast, MOF-5-$NH_2$, which possesses a larger lattice constant, formed a comparatively less dense hydrogen-bond network, resulting in poor sensing performance. Across the nanostrip series, proton conductivities were typically two orders of magnitude higher than those reported for most existing proton-conductive porous materials and MOF-based sensing layers [2,5,9,34-46] (Fig. 3e), indicating that the interfacial nanostrip geometry, rather than composition alone, is central to the performance gain.

The resulting sensing fibres were also mechanically and chemically robust (detailed in Supplementary Note 6). The nanostrip coating tolerated water-jet impacts exceeding ~30 m/s with <2% conductivity change, retained functionality after 1 week in ethanol and after a 3-h washing-machine cycle, and showed only ~10% conductivity drift after 1,000 repeated stretching and bending cycles (Fig. 3f, Supplementary Fig. 11 and Supplementary Video 1). These results show that the rapid protonic response is not restricted to delicate laboratory samples but is maintained in fibre devices compatible with practical handling and wearable use.

## Molecular origin of interfacial superprotonic conduction

To understand the origin of this nontrivial fast and strong protonic response, we compared proton transport in the bulk MOF and at the MOF interface using *ab initio* molecular dynamics simulations (Supplementary Note 13), supported by Arrhenius analysis of the experimentally measured conductivity (Fig. 3g). Consistent with established Grotthuss mechanisms in the bulk phase of UiO-66 [26,47], proton migration was strongly constrained at low hydration, and only became more favourable at higher water content when a denser hydrogen-bonded network formed within the cages (Extended Data Fig. 5a). In contrast, at the MOF interface, a fraction of the adsorbed water molecules remained bound near the metal-cluster sites, where they formed a flexible interfacial hydrogen-bond network, while the remaining hydronium-containing species stayed mobile and migrated continuously along the interface (Fig. 3h). Under low-humidity conditions, these mobile carriers were predominantly $H_3O^+$-type, whereas under higher humidity they evolved into larger Eigen-type ($H_9O_4^+$) hydrated species that also diffused readily within the interfacial water layer. These simulations therefore indicate that proton transport in the nanostrip fibres is dominated by interfacial vehicle-like conduction within a few-molecule-thick adsorbed water layer [27], rather than by pore-limited transport. This interpretation is further supported by the radial distribution functions and proton diffusion coefficients (Extended

Data Fig. 5c, d), which show a less confined hydration environment and substantially faster proton mobility at the interface than in the bulk MOF. Hence, the interfacial water layer shows the features of the molecular superionic water, where fast proton transport arises from a flexible hydrogen-bond network and facile carrier motion [25].

The interpretation of vehicle transport mechanism is supported by Arrhenius analysis, which gave activation energies $E_a$ above 0.4 eV for both UiO-66 and UiO-66-OH nanostrip fibres (Fig. 3g), distinct from lower reported values for UiO-66 powders[8,47]. Simulations of UiO-66-OH further showed that hydroxyl-functionalised linkers stabilise denser interfacial hydrogen-bonding networks than UiO-66, justifying the superior conductivity of the hydroxylated nanostrip fibres (Extended Data Fig. 5b).

Therefore, we attribute the rapid humidity sensing of the nanostrip fibres primarily to interfacial proton transport. Two additional structural features probably reinforce the kinetic behaviour: the ultrathin nanostrip geometry, which reduces slow pore-filling diffusion [48], and the exposed heterogeneous hydrophilic–hydrophobic surface [49], which may promote rapid water desorption during recovery (Extended Data Fig. 6). These characteristics together distinguish the templated nanostrip structure from the thicker coatings, whose transport is dominated by discontinuous intergranular contacts and slower diffusional equilibration.

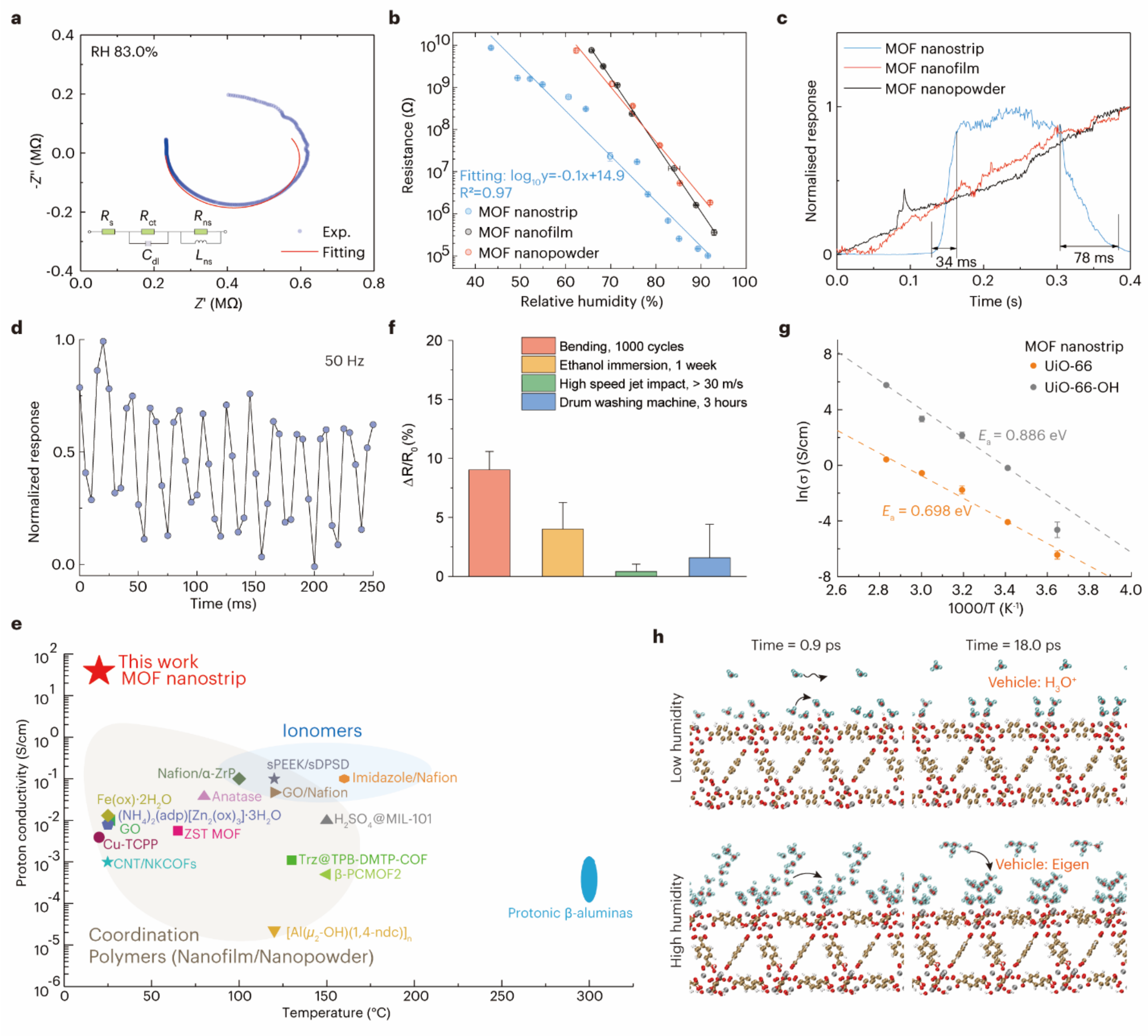

**Figure 3. Superprotonic conduction enabling rapid and robust humidity sensing**. (a) Nyquist plots of the impedance (imaginary Z'' versus real Z' parts of impedance) of MOF nanostrip-coated silk yarn under humid environments, showing low-resistance ionic conduction and an inductive feature associated with the nanostrip architecture. Equivalent circuits are overlaid to illustrate the fitting models, with symbols denoting charge transfer resistance ($R_{ct}$), series ionic resistance ($R_s$), double-layer capacitance ($C_{dl}$), inductance component ($L_{ns}$), and nanostrips-associated resistance ($R_{ns}$). (b) Humidity sensing performance across different relative humidity (RH) levels. (c) High-frequency impedance measurement under humidity modulation at 5 Hz using a mechanical chopper. (d) Same high-frequency measurement but at a humidity-modulation frequency of 50 Hz. (e) Comparison of proton conductivities of MOF nanostrips with previously reported proton-

conducting materials. The corresponding values from the literature are summarised in Supplementary Table 1. (f) Evaluation of sensor robustness and stability through resistance change assessment under 80% RH. (g) Arrhenius plot for proton conductivity of two MOF nanostrips at 80% RH. (h) Snapshots from the molecular dynamics simulations taken at t = 0.9 ps and at the end (18.0 ps), showing the proton transport along the MOF interface.

## Non-invasive respiratory monitoring with micro-thread

To demonstrate the practical utility in a clinically relevant and demanding setting, we integrated a single sensing micro-thread (with MOF nanostrip coating) into a standard medical nasal cannula and connected it to a small wireless circuit forming a complete respiratory monitoring system (Fig. 4a). The micro-threads (fibre sensors) possess several practical features that motivated this demonstration: they can be fabricated over large textile areas in a single batch, retain an imperceptible thread-like form factor, and can be positioned near the nasal nares by tethering or seamlessly integrated into an existing medical device (Extended Data Fig. 6). Such integration is clinically significant, as practical respiratory sensors must be lightweight, unobtrusive, and compatible with existing oxygen-delivery devices [50-52]. The ultra-thin thread-like form factor allows it to be mounted along the inner or outer wall of the cannula without obstructing airflow, while the wireless node enables wireless transmission of breathing signals to a tablet-based interface for real-time visualisation, respiratory rate (RR) extraction and alarm generation.

We first evaluated the micro-thread sensors in human volunteers. The sensor captured distinct breathing waveforms across different breathing patterns (Fig. 4b), enabled stable continuous RR monitoring over ~20 min (Fig. 4c), and distinguished inter-individual RR differences across 17 volunteers of different ages and sexes (Fig. 4d). A metronome provides regular timed cues that participants follow to breathe at a prescribed rate, providing a simple method for standardised paced-breathing assessments [53]. Using this approach, sensor-derived RR agreed closely with directly counted values (Fig. 4e), with negligible measurement bias (0.01 breaths $min^{-1}$; Supplementary Note 9). With only a small amount of MOF required and the coating remaining strongly bound to silk, the coated fibres (sensors) showed no detectable increase in cytotoxicity compared with pristine silk (Fig. 4f and Supplementary Note 7).

Next we tested the device under more challenging conditions relevant to clinical use. When integrated into a high-flow oxygen therapy system operating at up to 60 L $min^{-1}$, the sensor still resolved breathing signals across different supplied oxygen concentrations (Fig. 4g,h). Compared to spontaneous breathing under ambient conditions (Fig. 4b), the waveforms during high-flow oxygen therapy exhibited greater breath-to-breath variability in morphology. This likely reflects the dynamic mixing between the continuous oxygen flow and the exhaled air surrounding the sensor. The device also continued to capture respiratory activity during talking and walking with low motion artefacts (Fig. 4i). Together, these demonstrations show that our MOF nanostrip sensing thread can operate as an imperceptible respiratory monitor in a format already used for clinical oxygen delivery, providing accurate and continuous readout without adding disturbance to the current standard of care.

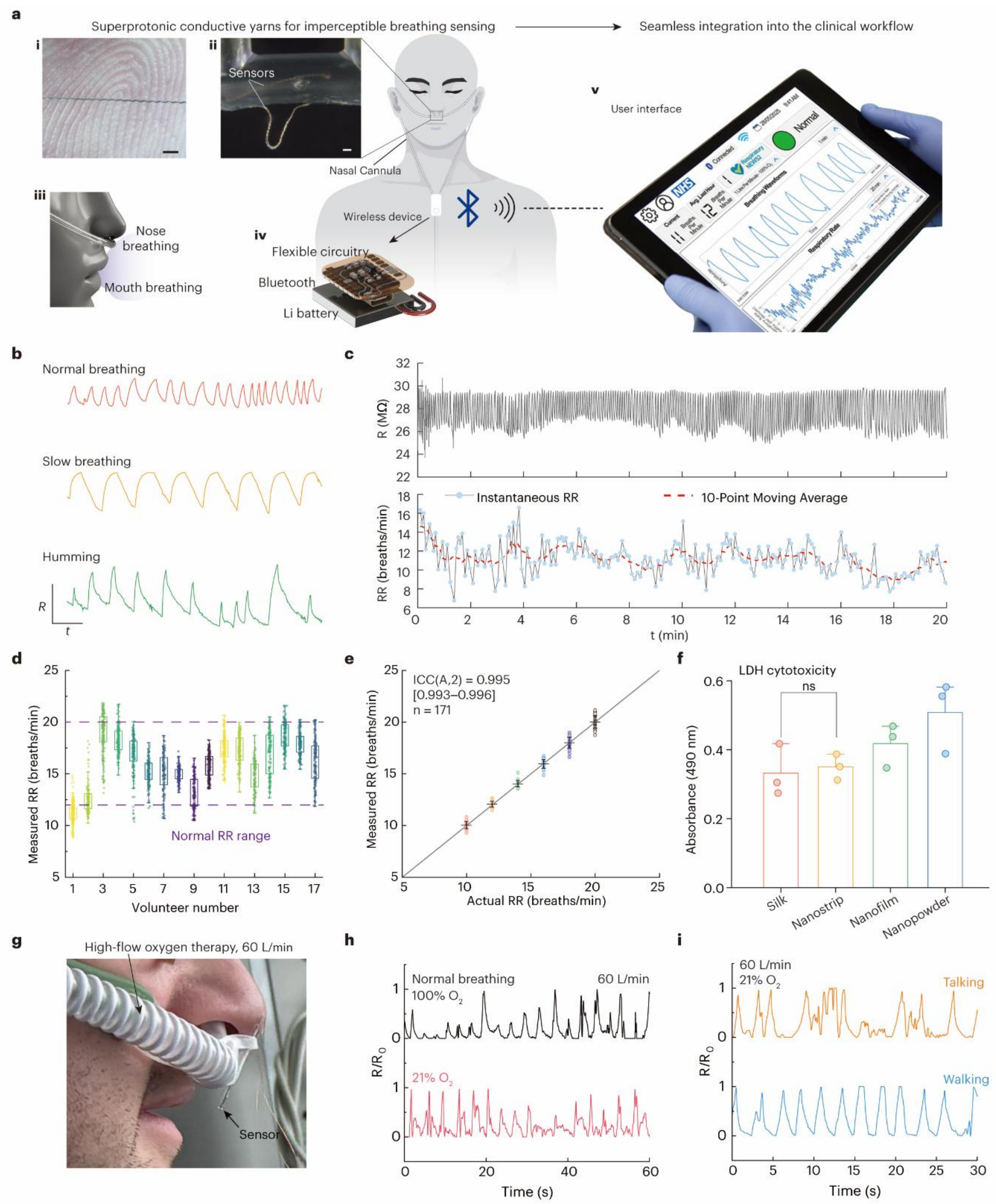

**Figure 4. Wearable respiratory monitoring using a micro-thread sensor.** (a) Platform overview of the respiratory monitoring system. (i) Photographs showing the small size of the sensing element and (ii) its integration inside and outside a nasal cannula. Scale bars: 1 mm. (iii) Side-view schematic of the face showing the sensorised nasal cannula, detecting local water vapour from both nose and mouth breathing. (iv) The wireless node,

including the flexible printed circuit, ESP32C3 Bluetooth chip and Li-ion battery. (v) Schematic of clinical use, in which respiratory data are transmitted wirelessly to a tablet-based monitoring interface. (b) Representative breathing waveforms measured from a healthy volunteer under different breathing patterns. The scale bar of time ($t$) and resistance ($R$) is 10 s and 10 MΩ, respectively. (c) Long-term (~20 min) respiratory measurement and the corresponding RR trace extracted from waveform peak detection. (d) Measured RR values from 17 volunteers of different ages and genders. (e) Comparison between sensor-derived RR and metronome-guided actual RR. Bland-Altman calculation was applied to assess the overall tracking capability. The overall intraclass correlation coefficient (ICC) is 0.995 (95% CI, 0.993–0.996); n=171 respiratory cycles from 3 volunteers. (f) LDH cytotoxicity assay of different silk-based samples. One-way ANOVA showed no significant difference among groups ($p$ = 0.058). (g) Photograph of a volunteer tested under a high-flow oxygen therapy system with a maximum flow rate of 60 L $min^{-1}$. (h) Respiratory signals measured under different supplied oxygen concentrations. (i) Respiratory signals measured during talking and walking, showing the ability of the sensor to capture breathing with low motion artefacts.

## Spatially resolved humidity mapping

To demonstrate that our sensors can be extended beyond single micro-threads to spatially resolved sensing, we fabricated textile humidity sensor arrays in both low-density and high-density formats (Fig. 5). Spatial mapping is a particularly demanding test because it requires not only local sensitivity at each sensing node, but also scalable patterning, parallel addressable readout and sufficient inter-pixel consistency to reconstruct meaningful humidity distributions. We first fabricated a 4 × 4 array by printing conductive silver ink onto the MOF nanostrip-coated textile (Fig. 5a,b) and positioned it beneath the nostrils to map breathing-humidity distributions (Fig. 5c). Each sensing unit consists of ~ 10 MOF nanostrip-coated fibres connected in parallel between two printed silver electrodes (bottom of Fig. 5b). The 4 × 4 array contains 16 such units sharing a common excitation electrode, and the 16 outputs are recorded in parallel through 16 independent analogue-input channels of the data acquisition card (readout circuit in Supplementary Fig. 20).

The array resolved the difference between symmetric breathing and partial occlusion of the left nostril, revealing clear spatial asymmetry in the measured humidity maps (Fig. 5d). This experiment shows that the nanostrip sensing textiles can operate not only as single-point RR monitoring, but also as addressable textile nodes for localised breathing-humidity mapping.

We then increased the spatial density by constructing an embroidered 16 × 16 array within an 8 mm × 8 mm textile area, corresponding to 256 pixels with a pixel pitch of ~500 μm (Fig. 5e; as schematics in Fig. 1g). Each sensing node was defined at the intersection of orthogonal row and column silver electrodes, with the crossed sensing fibres sandwiched between the electrodes to form a pixel. To read all pixels, the array shared 16 operational amplifiers through a multiplexer and a DAQ card (details in Supplementary Note 12). As the multiplexer switches each pixel on and off, it effectively applies a ~20 ms square-pulse excitation to that node. Because the individual sensing pixels have large signalling (RC) time constant, the node cannot fully discharge within the pulse, resulting in a characteristic transient (Fig. 5f). To calibrate the array, a deep learning pipeline (termed CalNet) was developed to extract features from these transient waveforms to predict humidity, followed by a super-resolution module for smoother map reconstruction (Fig. 5f,g). A static uniform humidity environment was applied using the saturated salt solution method and a scalar reference sensor was placed near the array. Each pixel used its own temporal encoder in CalNet, keeping pixel calibrations independent and preserving spatial variation. The calibrated array yielded representative humidity maps with a mean absolute error of 1.75% RH relative to the reference sensor and a maximum per-pixel Bland-Altman bias of approximately −0.8% RH (Fig. 5h–j). These results confirm that the fibre platform can be scaled to high-density textile arrays with consistent calibration under uniform humidity.

Next, the ability of the array to reconstruct non-uniform fields was validated against known patterns. When a wet tissue was placed on a laser-cut acrylic mask, the reconstructed maps resolved the pattern through moisture diffusion (Fig. 5k). Letters “UCL” (in the mask) are clearly visualised. Time-resolved measurements further visualised the lateral propagation of a humidity front when the wet tissue was positioned on one side of the array (Fig. 5l). These demonstrations show that the nanostrip design can be translated from imperceptible micro-thread sensors to printed or embroidered textile imagers for spatially resolved humidity sensing.

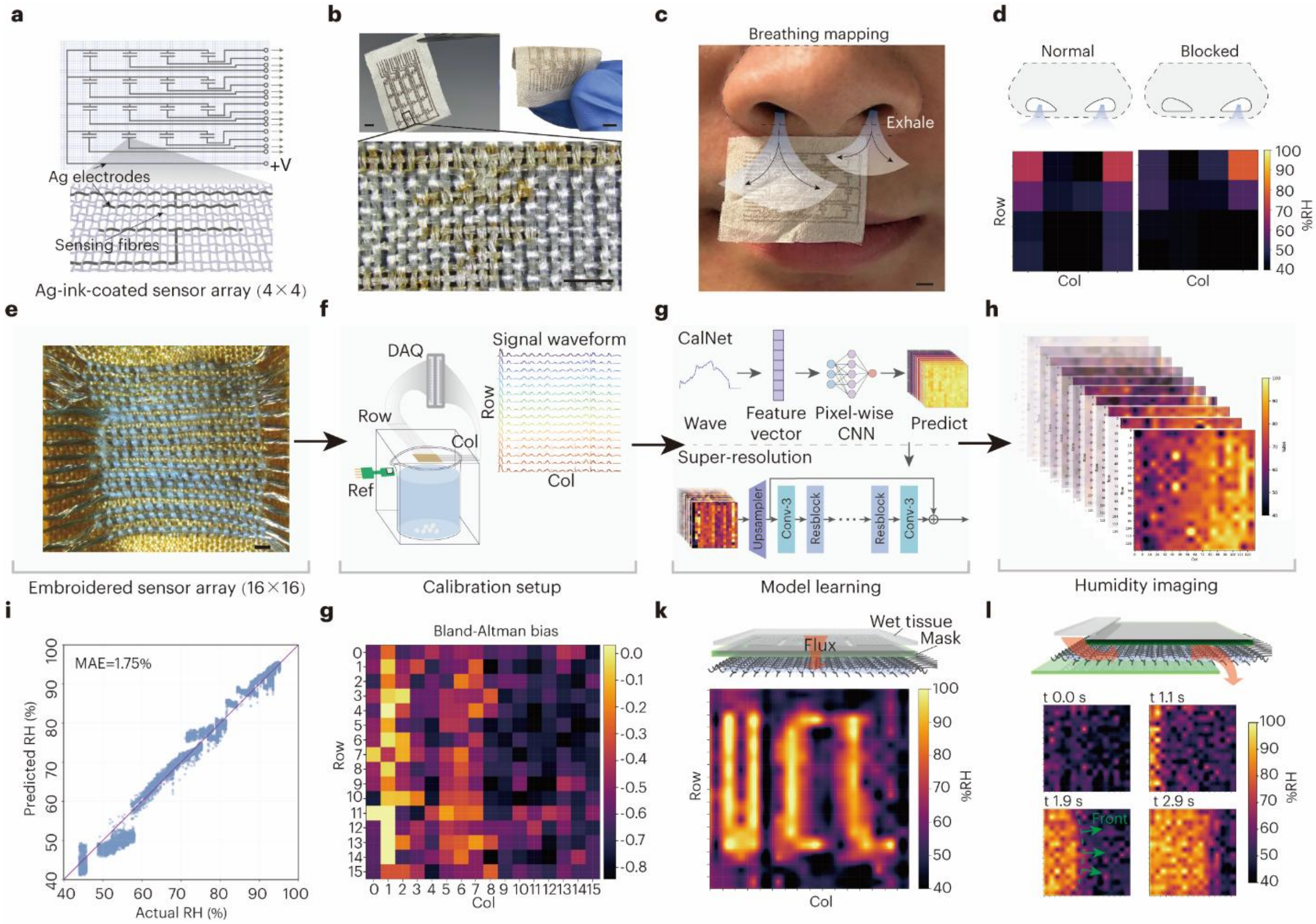

**Figure 5. Textile sensor arrays for spatial humidity mapping.**

(a) Schematics of the 4 × 4 humidity-sensing array fabricated by printing conductive silver ink onto the MOF nanostrip-coated conductive fabric. Top, the full array; bottom, an individual sensing unit. (b) Photographs of the array. Top, the array in flat and finger-folded states; bottom, optical micrograph of a single sensing unit. Scale bars: 4 mm (top) and 1 mm (bottom). (c) Photograph of the 4 × 4 array positioned beneath the nostrils for breathing-humidity mapping. (d) Spatial humidity maps measured with the 4 × 4 array during symmetric breathing (left) and after partial occlusion of the left nostril (right), showing the ability to resolve breathing asymmetry. (e) Optical micrograph of the embroidered 16 × 16 high-density sensor array. Scale bar: 1 mm. (f) Schematic of the calibration and data-acquisition setup for the 16 × 16 array, including a commercial humidity sensor, saturated salt solutions used to generate controlled humidity levels, and a DAQ-based readout circuit. Because the array was read out using 16 operational amplifiers and a multiplexer, each pixel produced transient pulse-like electrical responses rather than steady-state signals. (g) Schematic of the deep learning assisted calibration pipeline. The CalNet model extracts waveform features and performs pixel-wise prediction, and a super-resolution module is used to generate smoother humidity maps from the high-density array output. (h) Representative humidity maps reconstructed from the calibrated 16 × 16 array. (i) Comparison between actual and predicted RH values for the 16 × 16 array, with a mean absolute error (MAE) of 1.75% RH. (j) Per-pixel Bland-Altman bias map of the calibrated array, showing a maximum bias of ~−0.8% RH. (k) Humidity imaging of a wet tissue placed on a laser-cut acrylic mask, resolving the "UCL" pattern through moisture diffusion. (l) Time-resolved humidity mapping of a wet tissue placed on the left side of the array, showing lateral diffusion of the humidity front across the sensor. Owing to limitations of the readout circuit and the high impedance of the sensing elements, the present high-density array operates at around 1 frame per second.

## Outlook

The superprotonic fibre sensors provide continuous, real-time breathing waveforms, which could help detect clinical deterioration early, for example in patients recovering from major surgery. The imperceptible single-thread form is well suited to infants, older people and others for whom conventional monitors are uncomfortable. Moreover, the long-term waveforms they record could provide high-quality datasets for future predictive AI models. Because gases such as $NH_3$ and $CO_2$ are closely coupled to water and protons, the same interfacial proton-transport platform could be adapted with different MOFs to measure these gases. Among them, $CO_2$ is a widely accepted respiratory marker in clinical practice. For better wearability and sustainability, a further step would be to integrate sensing and wireless communication into a single fibre, reducing the use of readout chips. Having demonstrated spatial humidity imaging with woven arrays, we expect the same approach could be extended to map water-vapour loss and sweat evaporation across the skin, offering a possible route to monitor abnormal sweating function. Our template-directed interfacial growth strategy could be applied beyond MOF materials and humidity sensing function to other wearable sensors and imaging devices.

## Methods

### Chemicals

Zirconium(IV) chloride ($ZrCl_4$), Zirconium chloride octahydrate ($ZrCl_2 \cdot 8H_2O$), Zinc nitrate hexahydrate, terephthalic acid, 2-aminoterepthalic acid (ATPA), 2,5-Dihydroxyterephthalic acid (DHTPA), lithium bromide (LiBr), dimethyl formamide (DMF), chloroform, acetone, hydrochloric acid (HCl), and ethyl alcohol (≥99.5%) were purchased from Sigma Aldrich. 1-(3-Dimethylaminopropyl)-3-ethylcarbodiimide hydrochloride (EDAC) was purchased from Thermo Scientific. All the chemicals were used without further purification. 100% pure mulberry silk fabric was ordered from Beckford Silk. Distilled water was used for synthesize and testing. Commercial silver ink (DM-SIP-2005) was obtained from Dycotec.

## Characterisation

Surface morphologies and elemental compositions were characterised using SEM (GeminiSEM 300 system, Carl Zeiss, Germany) equipped with energy-dispersive X-ray spectroscopy (EDS). Before imaging, a thin gold film was sputter-coated onto the sample surfaces. Morphology and nanomechanical properties were also studied using an AFM (Bruker Multimode 8). Fabric samples were imaged using a ScanAsyst-Air probe in PeakForce Tapping mode to obtain high-resolution surface topography. To minimise movement during scanning, individual micro-threads were affixed to mica sheets using carbon tape. To measure the thickness of the MOF nanostrips, the nanostrip-coated silk samples were immersed in 9.3 M LiBr solution at 60 °C for 1 h to dissolve the silk fibroin template. The resulting solution was transferred into a Spectra/Por® 4 dialysis membrane (molecular weight cut-off ~3.5 kDa) and dialysed against deionised water for 48 h to remove LiBr. After dialysis, the solution was diluted tenfold with water to reduce concentration and facilitate the deposition of individual MOF nanostrips. The diluted suspension was then spin-coated onto mica sheets and imaged using a ScanAsyst-Air probe in PeakForce Tapping mode. FTIR spectra were collected using a spectrophotometer (Nicolet™ iS50, Thermo Fisher) in the wavenumber range of 400–4000 $cm^{-1}$. Optical images of the humidity fabric sensors and human hair were captured using a digital microscope (Keyence, VHX-7000). Contact angles were measured using a custom-built goniometer setup [23]. The thermal properties of the samples were evaluated by TGA. Approximately 5 mg of each sample was placed in a platinum crucible and heated from 30 °C to 800 °C at a ramp rate of 20 °C/min under a nitrogen atmosphere (Discovery, TA Instruments, USA). The percentage of mass loss was calculated from the resulting TGA curves. XRD patterns were recorded using a Malvern Panalytical Aeris X-ray diffractometer. Tensile tests and cyclic bending durability tests of the fibres were performed using a universal testing machine (Instron, model 5969) at a constant speed of 1 mm/min. The initial toe region of some curves, arising from initial slack take-up, was corrected using the tangent-intersection method (ASTM D638).

## Device fabrication

**Fabrication of MOF nanostrips on silk.** MOF nanostrips were fabricated *via* a layer-by-layer growth strategy on a silk template. Prior to MOF deposition, the silk was cleaned sequentially with acetone and DMF to remove surface contaminants. The pre-cleaned silk was then immersed in a DMF solution containing the MOF organic linker (25 mM) and 1 wt% coupling agent EDAC. This pre-treatment was carried out in a sealed glass vial at 80 °C for 6 hrs to facilitate β-sheet crystallisation of the silk fibroin and enable covalent grafting of the linker onto the surface proteins. Following this step, the silk was rinsed with DMF and subsequently immersed in a 25 mM DMF solution of metal salt at 120 °C for 10 minutes. After another DMF rinse, the silk was re-immersed in the linker solution for 10 minutes, completing one cycle of MOF growth. Finally, the MOF-functionalised fabric was thoroughly washed with DMF and chloroform to remove residual unreacted metal salts and linkers, followed by drying under vacuum at 100 °C overnight. UiO-66 was synthesised using terephthalic acid and $ZrCl_2·8H_2O$ as organic linker and metal salt, respectively. Hydrophilic MOFs, UiO-66-$NH_2$ and UiO-66-OH, used ATPA and DHTPA as different linkers. MOF-5-$NH_2$ was composed of zinc nitrate hexahydrate and ATPA.

**Control sample preparation.** Solvothermal growth was applied to fabricate MOF nanofilm on silk. Briefly, the silk fabrics were washed with acetone and DMF before coating. Silk was then immersed in the mixture of DMF solution of MOF linker (25 mM) and $ZrCl_2.8H_2O$ (25 mM) in a tightly closed glass bottle overnight at 120 °C overnight. The fabric was thoroughly washed in DMF and then chloroform. The surface was vacuum dried overnight at 100 °C. MOF nanopowder coatings were fabricated by spraying MOF nanoparticles dispersion on silk. ATPA (67 mg), $ZrCl_4$ (67 mg), and HCl (1 mL) were dissolved in 15 mL of DMF followed by heating overnight at 120 °C. The light-yellow precipitates were centrifuged and washed thrice with DMF and acetone. MOF nanoparticle suspensions were spray-coated onto silk fabrics using an airbrush (Iwata Eclipse, ECL2000) operated at 2.5 bar pressure, followed by drying overnight at 100 °C under vacuum.

**Fabrication of micro-thread sensor.** The MOF-coated silk fibres can be configured into complete sensing devices by either painting silver ink electrodes or knotting them with conductive wires (detailed in Supplementary Note 8 and Supplementary Figure 11). The sensing thread is designed for versatile integration with standard medical interfaces. Specifically, the MOF-coated fibre can be easily inserted into or externally adhered to commercial breathing circuits (e.g., Intersurgical nasal cannulae) without obstructing the airflow (Fig. 4a). The sensing unit was subsequently connected with a flexible polyimide-based circuit board fabricated via direct ink writing printing and the circuit is shown in Supplementary Fig. 17. The circuit was integrated with an ESP32-C3 module (Seeed Studio) for wireless data transmission and a Li-ion polymer battery (190 mAh) as the power source (see Fig. 4a). The assembled system functioned as a wearable wireless respiratory monitoring device and was employed for human subject testing.

**Fabrication of humidity sensor array.** Humidity sensor arrays were fabricated directly from the MOF nanostrip-coated conductive textile by exploiting the intrinsic yarn geometry and open pore structure of the fabric. For the 4 × 4 array, conductive silver ink (DM-SIP-2005) was manually patterned onto the textile using a fine dispensing needle to define orthogonal interconnects along selected yarn pathways. The exposed MOF-coated textile between the silver-coated lines served as the humidity-sensitive pixels. For the 16 × 16 high-density array, conductive rows and columns were introduced by embroidery, using the inter-yarn pores as accessible pathways for stitching without disrupting the textile architecture. Each stitched junction defined an individual sensing unit. In both cases, the woven substrate acted simultaneously as the mechanical support, electrode-routing template and sensing scaffold, enabling low-density printed arrays and high-density embroidered arrays to be constructed from the same textile platform while preserving flexibility and conformability.

## Electric performance measurement

An impedance analyser (Hioki IM3570) was used to perform measurements of electrochemical impedance spectroscopy and proton conductivity on fabric sensors. The distance between the two microprobe electrodes was controlled at ~300 μm. Sensors were studied inside an environmental chamber (LTCL400, TAS UK Ltd.) with controllable temperature and humidity. Impedance spectra were acquired across a frequency range from 50 to $5 \times 10^6$ Hz. For high-frequency response measurements, the analyser was operated in continuous single-frequency mode and the excitation frequency was fixed at 1 kHz. Humidity was modulated at 5 Hz and 50 Hz using a mechanical chopper (Supplementary Note 3). Steady-state resistance was recorded using an RS PRO RSDM3055 Bench Digital Multimeter. For micro-thread sensors, real-time resistance changes during respiratory monitoring were captured by a bespoke wireless node (see details in Supplementary Note 10). To achieve high-throughput spatial sensing, the 16 × 16 sensor array was interfaced with a custom-built readout system. The system includes a data acquisition card (National Instruments USB-6341) and operational amplifiers (ADA4530-1), which provided the ultra-low input bias current necessary for the stable measurement of high-impedance sensing elements. Detailed configurations of the circuit architecture are provided in Supplementary Note 12.

## Human participant studies

**Ethics and Institutional Approvals.** The human pilot study was conducted in accordance with the Declaration of Helsinki and approved by the UCL Life and Medical Sciences Research Ethics Committee (Project ID: 1269; Sponsored by UCL ROBUST DE-RISK, EDGE ID: 180997). Informed consent was obtained from all 17 participants.

**Respiratory Monitoring and Pre-clinical Validation.** To evaluate the MOF nanostrip-coated micro-thread sensor, the device was integrated with a nasal cannula for respiratory sampling in a climate-controlled room (20°C, 40% RH). The study comprised two phases: (i) ambient air monitoring at the UCL laboratory under non-oxygen-supplemented conditions, and (ii) clinical validation at UCL Hospital using a hospital-standard high-flow oxygen therapy system to assess sensor resilience under high-humidity medical settings. Detailed experimental protocols, participant demographics, and equipment specifications are provided in Supplementary Notes 9-11.

## Statistical analysis

**Regression fitting.** Linear regression was applied to the data in Fig. 3b and 3g using OriginPro 2025b. Goodness of fit is reported as the coefficient of determination ($R^2$).

**Agreement analysis.** Agreement between sensor-derived and reference was assessed by Bland–Altman analysis using R (version 4.6.), and the ICC was used to quantify overall tracking capability (Fig. 4e). The ICC was 0.995 (95% CI, 0.993–0.996), calculated as ICC(A,2) (two-way random effects, absolute agreement, average measures). This analysis was based on $n = 171$ respiratory cycles from 3 volunteers.

**ANOVA.** A one-way ANOVA compared LDH release among silk-based samples (Fig. 4f; 3 independent samples per group). The analyses were performed in R (version 4.6.). No significant difference was found among groups ($p = 0.058$). Data are presented as mean ± s.d.

### Acknowledgements

We acknowledge funding from InspiringFuture European Research Council (ERC) Consolidator Fellowship selected by the ERC, funded by UKRI Horizon Europe Guarantee (EP/X023974/1), Royal Society Wolfson Fellowship (RSWF\R3\193013), and support from the UCL Hawkes Institute.

### Contributions

J.Z. and M.K.T. conceived the idea. J.Z., A.S. performed all measurements, V.S. contributed to chemical characterisation, H.W. performed cytotoxicity tests, J.Z. undertook molecular simulations and A.S. designed the wireless node software. Human volunteer tests were performed by J.Z., A.S. and R.T. R.T., L.B.L. and M.K.T. provided supervision and guidance. All authors contributed to results analyses and interpretation. J.Z. wrote first draft with support from A.S., following which all authors made revisions and edits to finalise the paper.

**Extended Data Fig. 1 Schematic and SEM characterisation of various MOF-functionalised fibres.** (a) Schematic illustration of conventional MOF structure used as controls, contrast with our approach: a solvothermally grown MOF nanofilm and a spray-coated MOF nanopowder layer. (b) Surface morphologies and structure of diverse silk-based fibres. SEM images of pristine silk fibre, and a series of MOF-functionalised fibres, including MOF nanostrip coating synthesised *via* 3 cycles of layer-by-layer (LbL) assembly, MOF nanofilm coatings prepared *via* 5 and 10 LbL cycles, respectively, and MOF nanofilm and MOF nanopowder coatings.

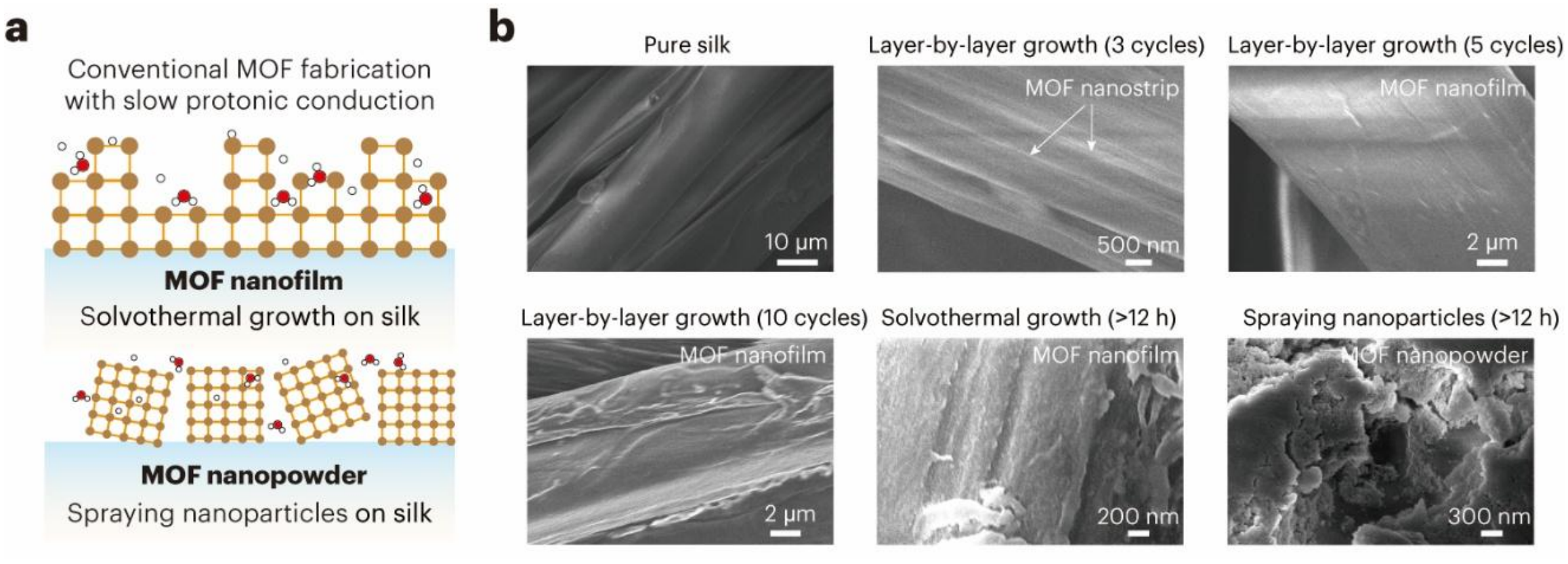

**Extended Data Fig. 2 Surface topography analysis.** Surface morphologies of diverse fibres imaged using AFM, including (a) pristine silk fibre, and (b,c) fibres modified via LbL assembly (various cycles), (d) solvothermal growth, and (e) nanoparticle spray-coating. Scale bar: 100 nm. (f) Surface roughness analysis derived from AFM topography scans across different structures. Each bar represents the mean arithmetic roughness (Ra), with individual data points (dots) representing measurements from three independently prepared batches. Error bars indicate the standard deviation.

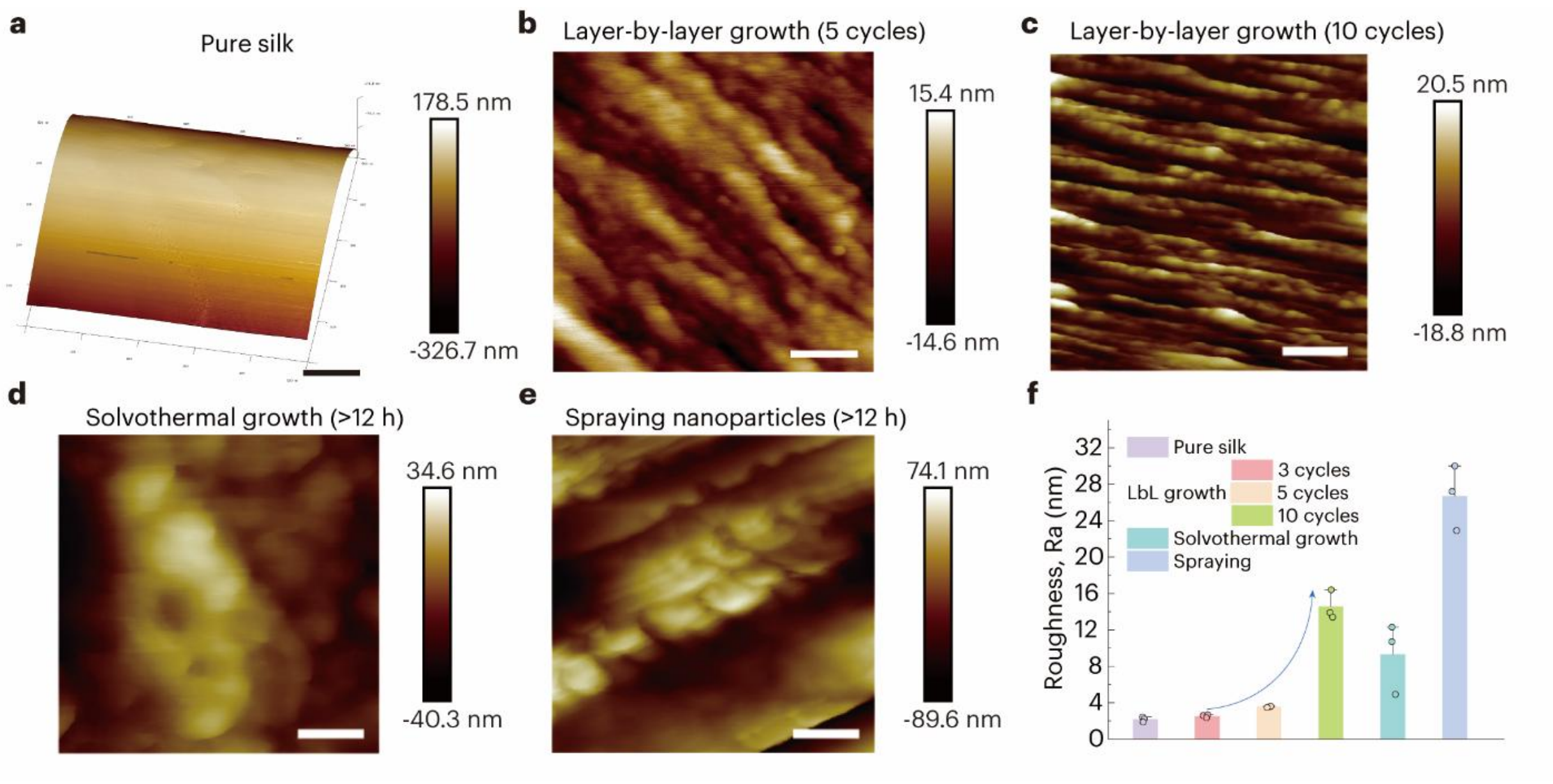

**Extended Data Fig. 3. Electrochemical impedance spectroscopy (EIS) and proton conductivity.** (a) Nyquist plots of the impedance (imaginary part Z'' versus real part Z') of MOF nanostrip-coated silk yarn under humid environments. Overlaid equivalent circuits represent the fitting models, where $R_s$, $R_{ct}$, $C_{dl}$, $Z_w$, $L_{ns}$, and $R_{ns}$ denote series resistance, charge transfer resistance, double-layer capacitance, Warburg impedance, inductance, and nanostrip-associated resistance, respectively. (b) Humidity response of MOF sensors fabricated through LbL assembly over different number of cycles. (c) Humidity response and (d) Nyquist plots of nanostrips synthesized with different MOF types. (e) Proton conductivity of the MOF nanostrip materials.

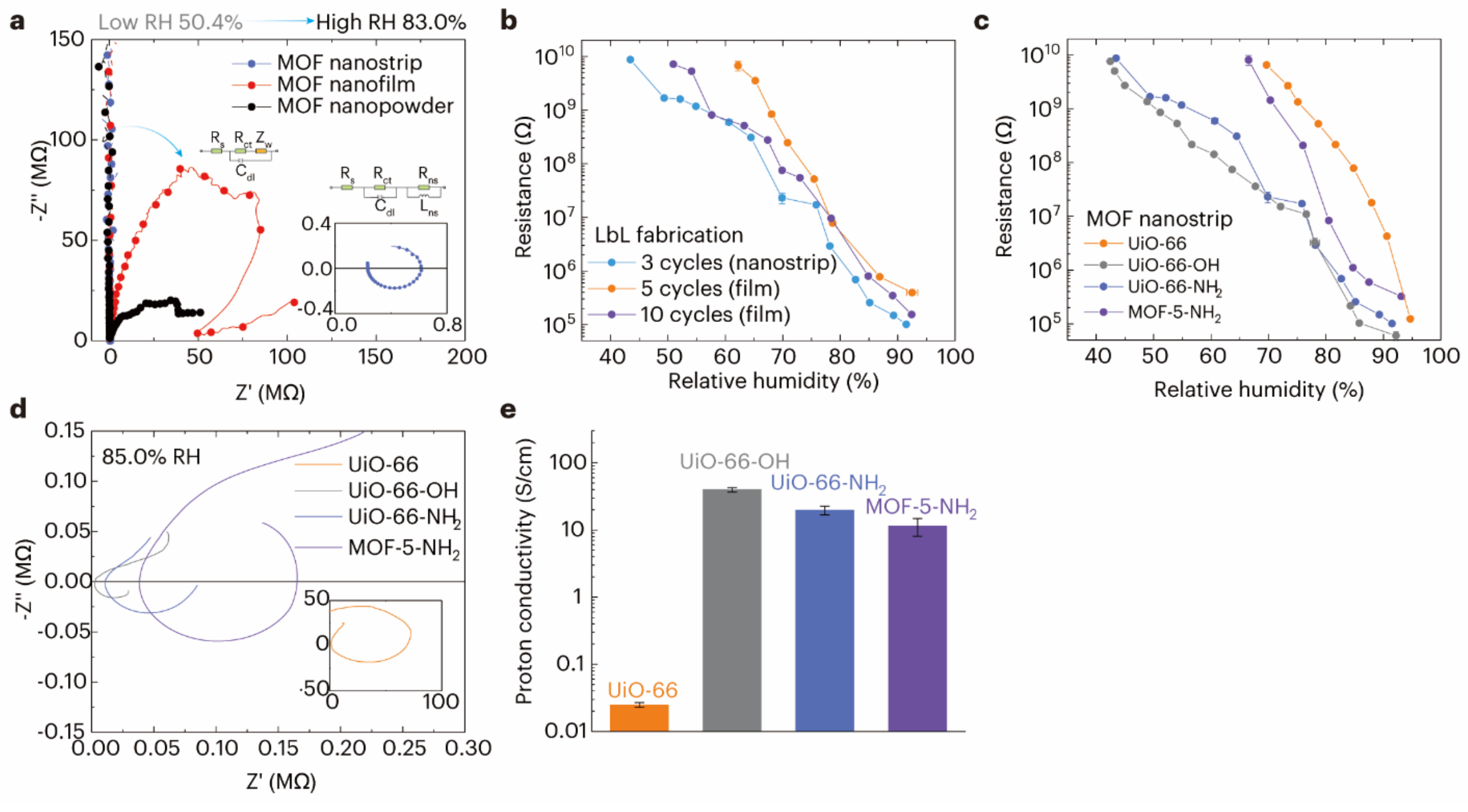

**Extended Data Fig. 4. Molecular simulations of proton conduction.** (a) Snapshots from the AIMD simulation taken at t = 1.8 ps and at the end (18 ps), showing the water and proton confined in the tetrahedral and octahedral cages of bulk-phase MOF. The crystal unit from the [110] projection is marked by gray lines. At high humidity cases, the water-mediated pathway formed in cages, which facilitated Grotthuss hopping. (b) Snapshots from the AIMD simulation of UiO-66-OH, showing the proton transport on the interface of MOF with denser hydrogen network. (c) Radial distribution function of oxygen in water and (d) proton diffusion coefficient.

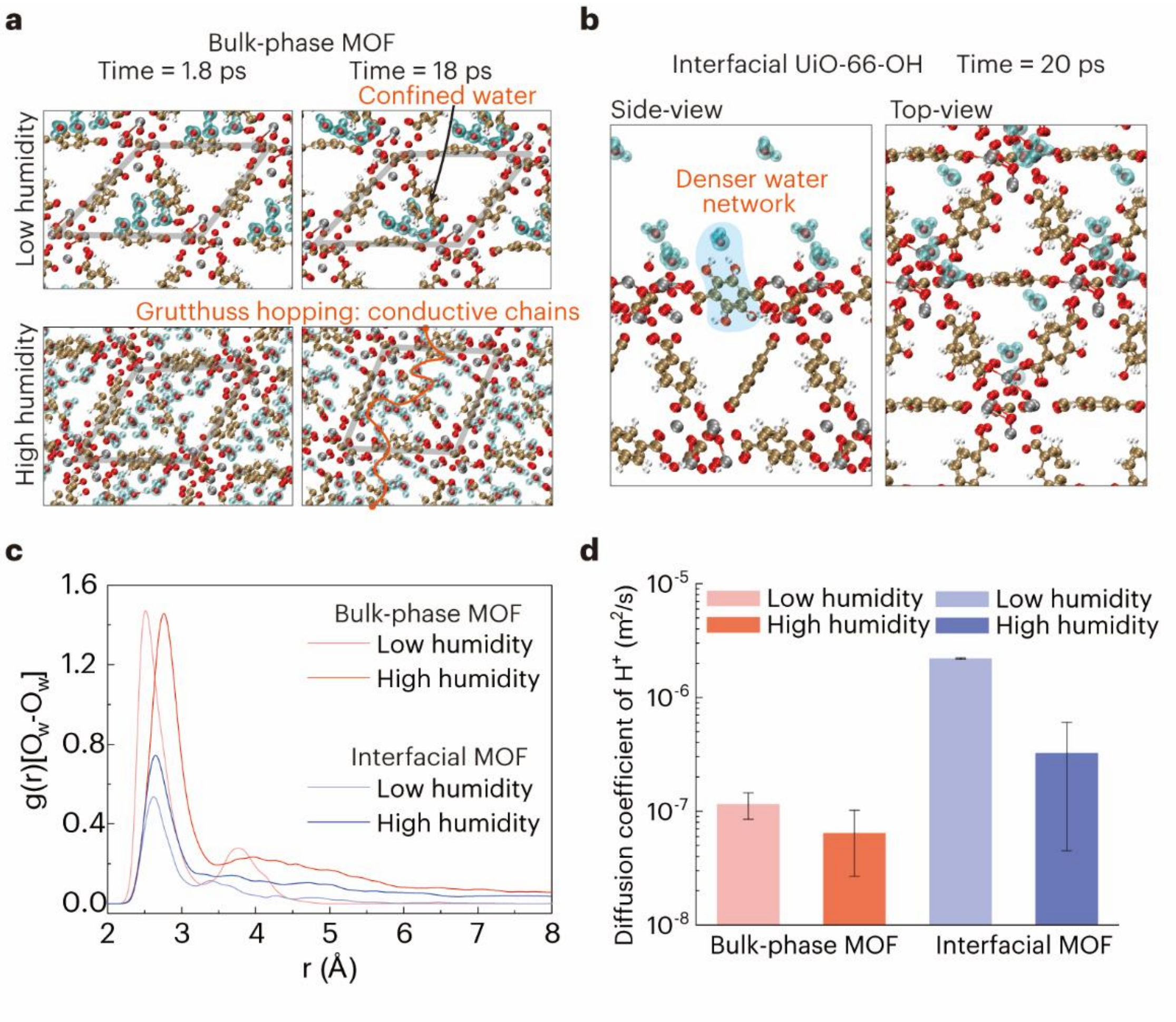

**Extended Data Fig. 5. Proposed sensing mechanism and impedance logic.** (a) Schematic illustration of the superprotonic sensing mechanism, depicting the synergistic proton transport pathways within the MOF nanostrip system. (b) Structural impedance models. Schematic diagrams comparing proton transport networks.

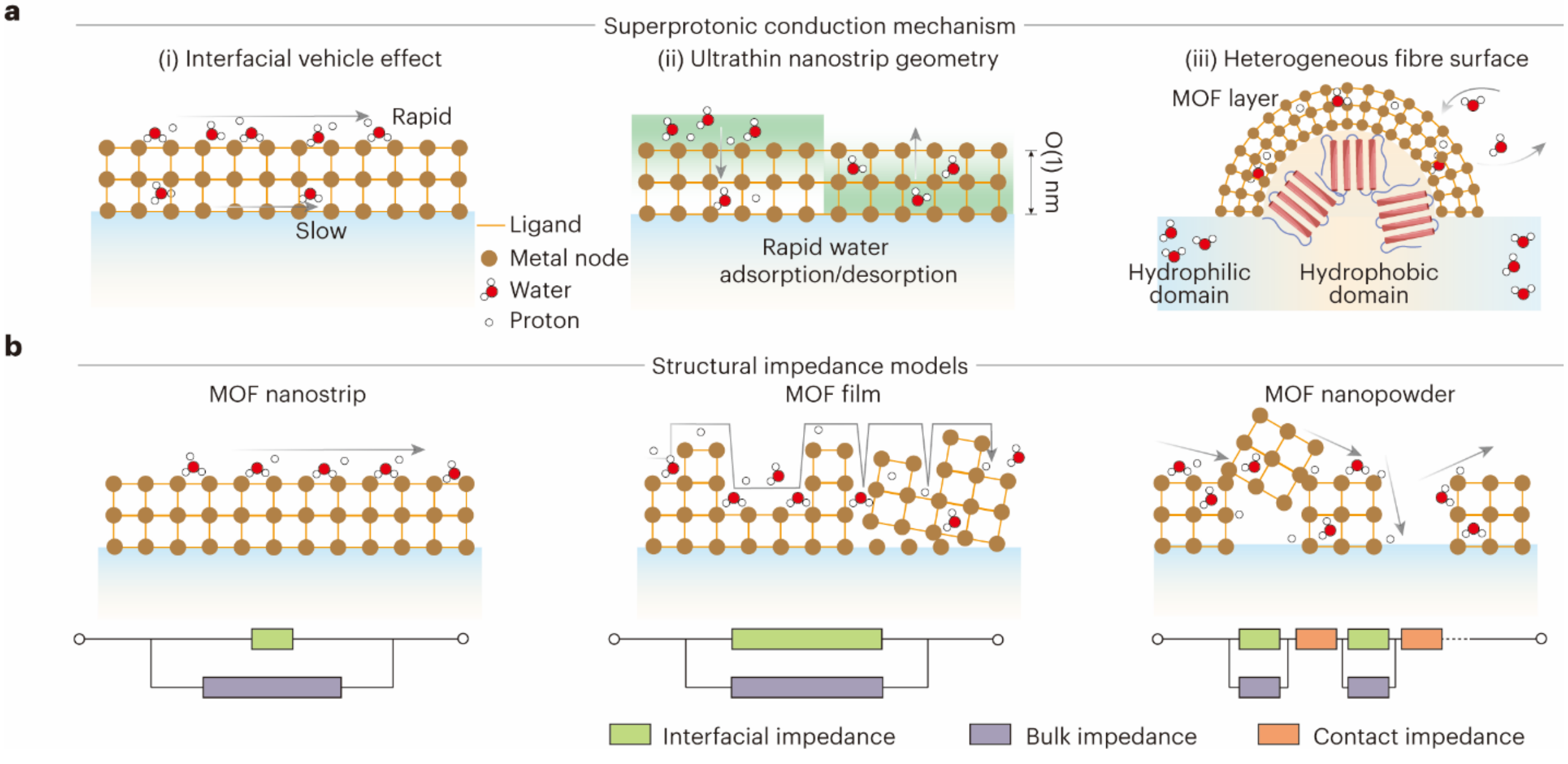

**Extended Data Fig. 6. Practical advantages of the MOF-functionalized fibres for biomedical sensing.** (a) Scalable fabrication. Photograph of a large-scale silk textile coated with MOF nanostrips in a single batch, demonstrating the potential for mass production. Scale bar: 10 cm. (b) Structural imperceptibility. Comparison of the feature sizes of biological skin structures [54] with the dimensions of the MOF-functionalised sensing units (right). The micro-thread sensor exhibits a fibre width of ~10 μm and a yarn width of ~100 μm, ensuring minimal physical presence on the skin. (c) Clinical adaptability. Photographs demonstrating the versatile integration of the micro-thread sensors, including direct positioning near the nasal vestibule and seamless sewing into the interior of a standard face mask. Scale bar: 20 mm.

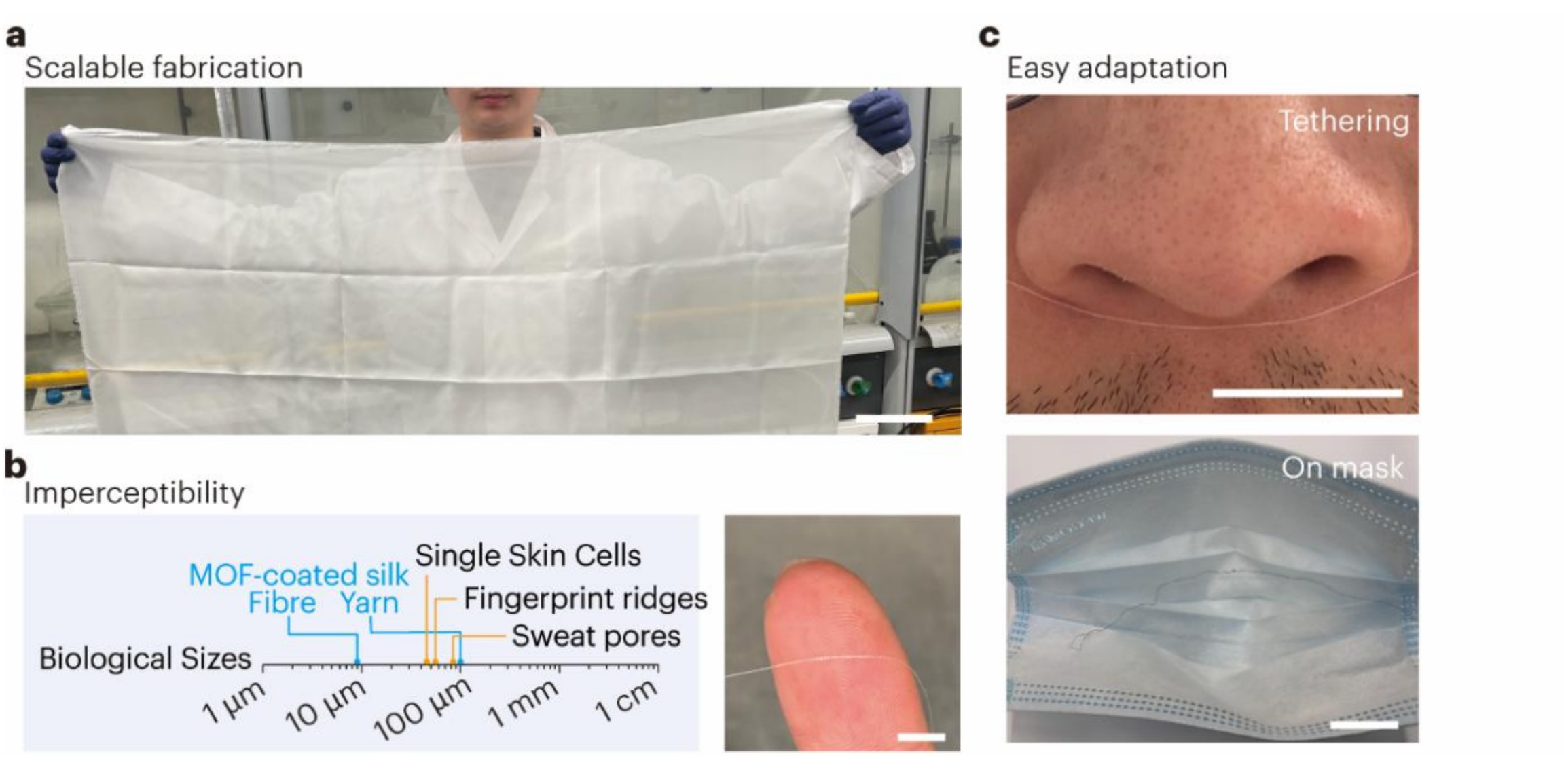

# Supplementary information

# Silk-templated Nanostrips for Superprotonic Fibre Interfaces

Jianhui Zhang[1,2], Ahmed Salem[1,2,3], Robert Tidswell[4], Haowei Wang[5,6], Vikramjeet Singh[7], Laurence B. Lovat[2,4], Manish K. Tiwari[1,2,7]*

[1]Nanoengineered Systems Laboratory, UCL Mechanical Engineering, University College London, London, WC1E 7JE, UK.

[2]Hawkes Institute, University College London, London, W1W 7TS, UK.

[3]UCL Medical Physics and Biomedical Engineering, University College London, London, WC1E 7JE, UK.

[4]Department of Critical Care, University College London Hospitals NHS, NW1 2BU, UK.

[5]Centre for Precision Healthcare, UCL Division of Medicine, University College London, London, WC1E 6JF, UK.

[6]Division of Biomaterials and Tissue Engineering, Royal Free Hospital, University College London, London, NW3 2PF, UK.

[7]Manufacturing Futures Lab, Mechanical Engineering, University College London, London, E20 2AE, UK.

***Corresponding author. Email: m.tiwari@ucl.ac.uk**

**This file includes:**

* Corresponding author, email: m.tiwari@ucl.ac.uk, phone: +44 20 3108 1056

Supplementary Notes 1 to 13
Supplementary Figures 1 to 21
Supplementary Tables 2
Supplementary movies 1 to 3
Supplementary References

## Supplementary Note 1. Theoretical models of fibre wettability

To evaluate the wetting behaviour of MOF nanostrip-coated silk fibres, we employed modified Wenzel and Cassie–Baxter models tailored for cylindrical geometries and hierarchical surface structures, as previously reported by [1].

For smooth silk fibres, the apparent contact angle ($\theta^*$) is estimated using a cylindrical Wenzel model:

$$\cos\theta^* = D^*_{fibre}\cos\theta_Y \quad (1)$$

Here, $\theta_Y$ is the Young's contact angle on flat silk, and $D^*_{fibre}$ is the roughness ratio specific to the cylindrical fibre system (see Supplementary Fig. 1), defined as:

$$D^*_{fibre} = \frac{R+D}{R} \quad (2)$$

where 2D is the spacing between fibres and 2R is the fibre diameter. As shown in Supplementary Fig. 2a, the $D^*_{fibre}$ of our silk fabric is 1.92.

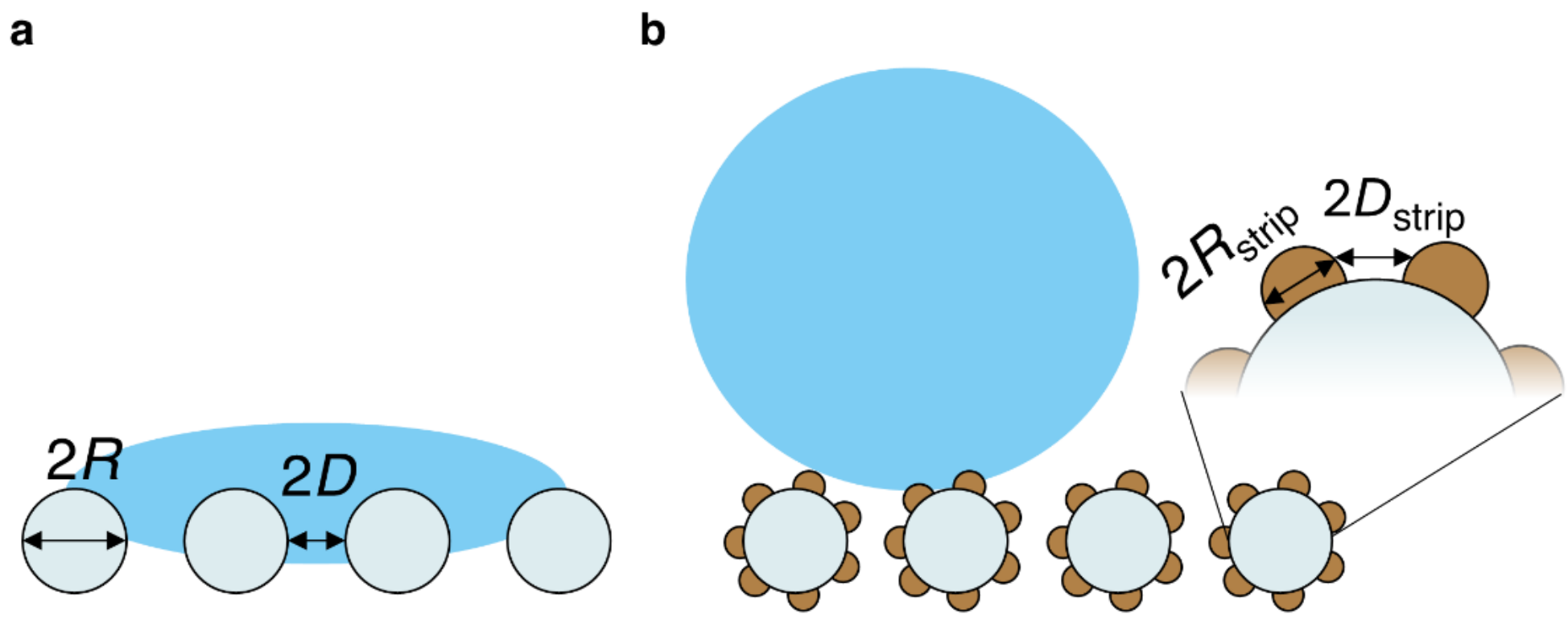

**Supplementary Figures 1 Schematics illustrating the roughness ratio parameters**. (a) Characteristic geometric parameters for plain woven fabrics. (b) Modified model incorporating nanostrips on individual fibres to account for hierarchical surface features.

For MOF nanostrip-coated fibres with hierarchical surface features, the wetting state transitions to a modified Cassie–Baxter regime, and the apparent contact angle is expressed as:

$$\cos\theta^* = -1 + \frac{1}{D^*_{fibre}}\left[\sin\theta^*_{strip} + (\pi - \theta^*_{strip})\cos\theta^*_{strip}\right] \quad (3)$$

where $\theta^*_{strip}$ is the contact angle on a smooth surface coated with nanostrips, which can be written as:

$$\cos\theta^*_{strip} = -1 + \frac{1}{D^*_{strip}}[\sin\theta_Y + (\pi - \theta_Y)\cos\theta_Y] \quad (4)$$

where $D^*_{strip}$ denotes the roughness ratio associated with the nanostrip structures protruding from the surface of the cylindrical fibre, and can be expressed as,

$$D^*_{strip} = 1 + \frac{D_{strip}}{R_{strip}} \quad (5)$$

where is $2D_{strip}$ the inter-strip spacing and $2R_{strip}$ is the strip diameter. Based on AFM analysis, the roughness ratio $D^*_{strip}$ was determined to be approximately 2.17 for the pre-treated silk surface and 1.81 for the MOF nanostrip-coated surface (Supplementary Fig. 2b and 2d).

The geometrical parameters extracted from the fibre surface were used to calculate the apparent contact angles, as shown in Supplementary Fig. 2e. The Young's contact angle of silk fibroin was taken as 74°, based on previously reported values measured on smooth regenerated silk fibroin film [2]. For UiO-66-$NH_2$, a Young's contact angle of ~62° was obtained from our measurements on a five-layer film assembled *via* a layer-by-layer strategy on glass substrates, following the same fabrication procedure as described in reference [3]. After pre-treatment with MOF linkers and metal nodes, the advancing contact angle on silk fibres increased to ~141° (Supplementary Fig. 2c), in close agreement with theoretical predictions (see Supplementary Fig. 2e). This indicates that the formation of nanostrip-like surface structures induces a transition to a non-wetting state, regardless of whether a MOF coating is present.

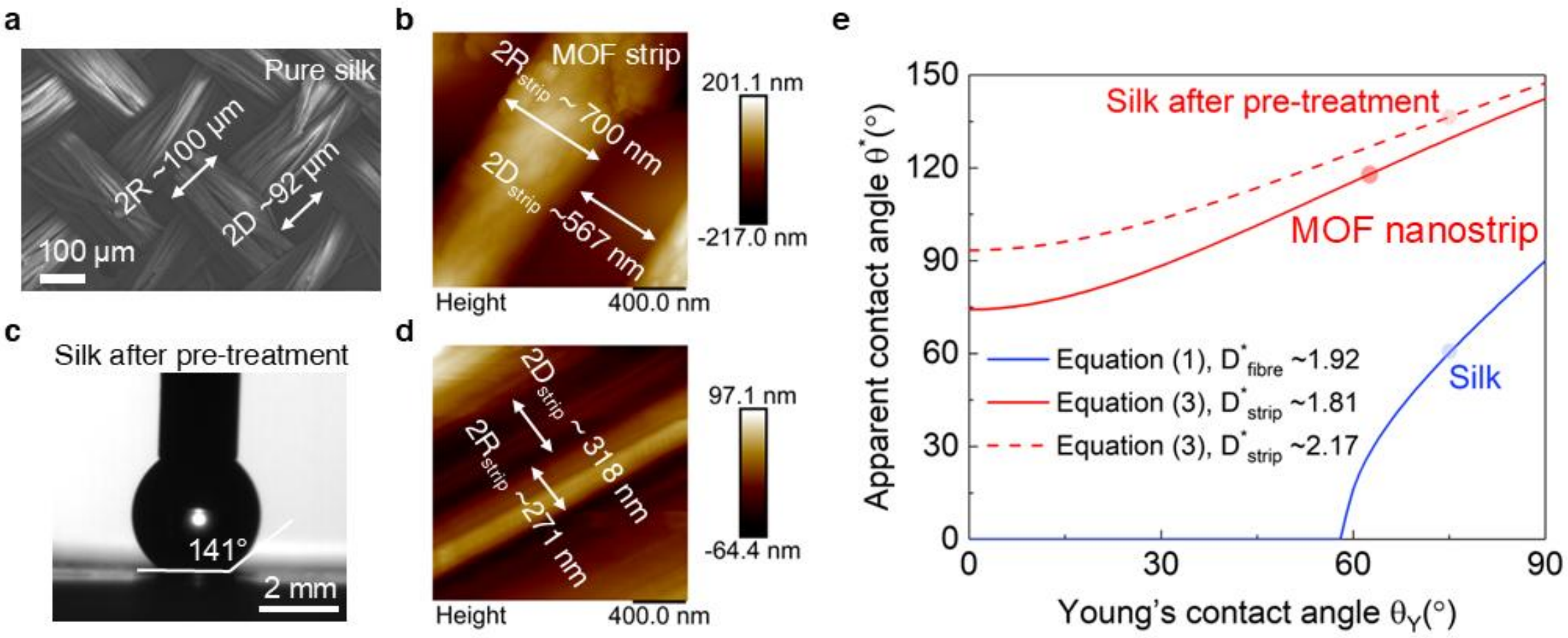

**Supplementary Figures 2 Geometrical characterisation and wettability prediction of fibre samples.** (a) SEM image of a pristine silk fibre, showing the smooth cylindrical morphology. (b) AFM topography of MOF nanostrips on the silk, revealing hierarchical surface roughness. (c) Advancing water contact angle on silk fabric after pre-treatment. (d) AFM image of a silk fibre surface after pre-treatment. (e) Predicted apparent contact angles for silk, silk after pre-treatment, and NP-MOF-coated fibres based on modified Wenzel and Cassie–Baxter models incorporating fibre curvature and surface structural parameters.

## Supplementary Note 2. Characterisation of MOF nanostrip-coated silk

Additional structural, chemical, mechanical and thermal characterisations were performed to complement the main analysis of the MOF nanostrip-coated silk fibres. These measurements further support the selective growth of the MOF nanostrips on the silk template, the compositional identity of the coating, and the mechanical and thermal characteristics of the resulting sensing fibres.

PeakForce quantitative nanomechanical maps using atomic force microscopy (AFM; Bruker Multimode 8) provided additional evidence for the heterogeneous surface organisation of the MOF nanostrip-coated silk fibres. In particular, the adhesion and dissipation channels clearly distinguished the MOF-coated regions from the intervening protein-rich amorphous regions (Supplementary Fig. 3). The amorphous silk regions exhibited higher adhesion due to their softer, more compliant nature, whereas the nanostrip-coated domains showed lower adhesion and greater surface roughness.

A representative PeakForce image (Supplementary Fig. 3) further revealed that, after three layer-by-layer assembly cycles, MOF nanostrips preferentially formed along nanofibrillar regions associated with the more crystalline domains of silk, while the intervening amorphous regions contained only sparse and disconnected MOF islands. This observation suggests that MOF growth is favoured on the crystalline nanofibrillar template and remains locally restricted in the amorphous regions during the early stages of assembly.

With additional growth cycles, however, this selective morphology was progressively lost. In particular, after five assembly cycles, SEM images (Extended Data Fig. 1b) showed that the distinct strip-like surface texture largely disappeared, consistent with overgrowth of the initially isolated MOF islands until they became laterally connected. These results support a growth mechanism in which the multiscale silk surface first directs anisotropic and selective nanostrip formation, but excessive layer-by-layer assembly eventually leads to coalescence and loss of the characteristic nanostrip architecture.

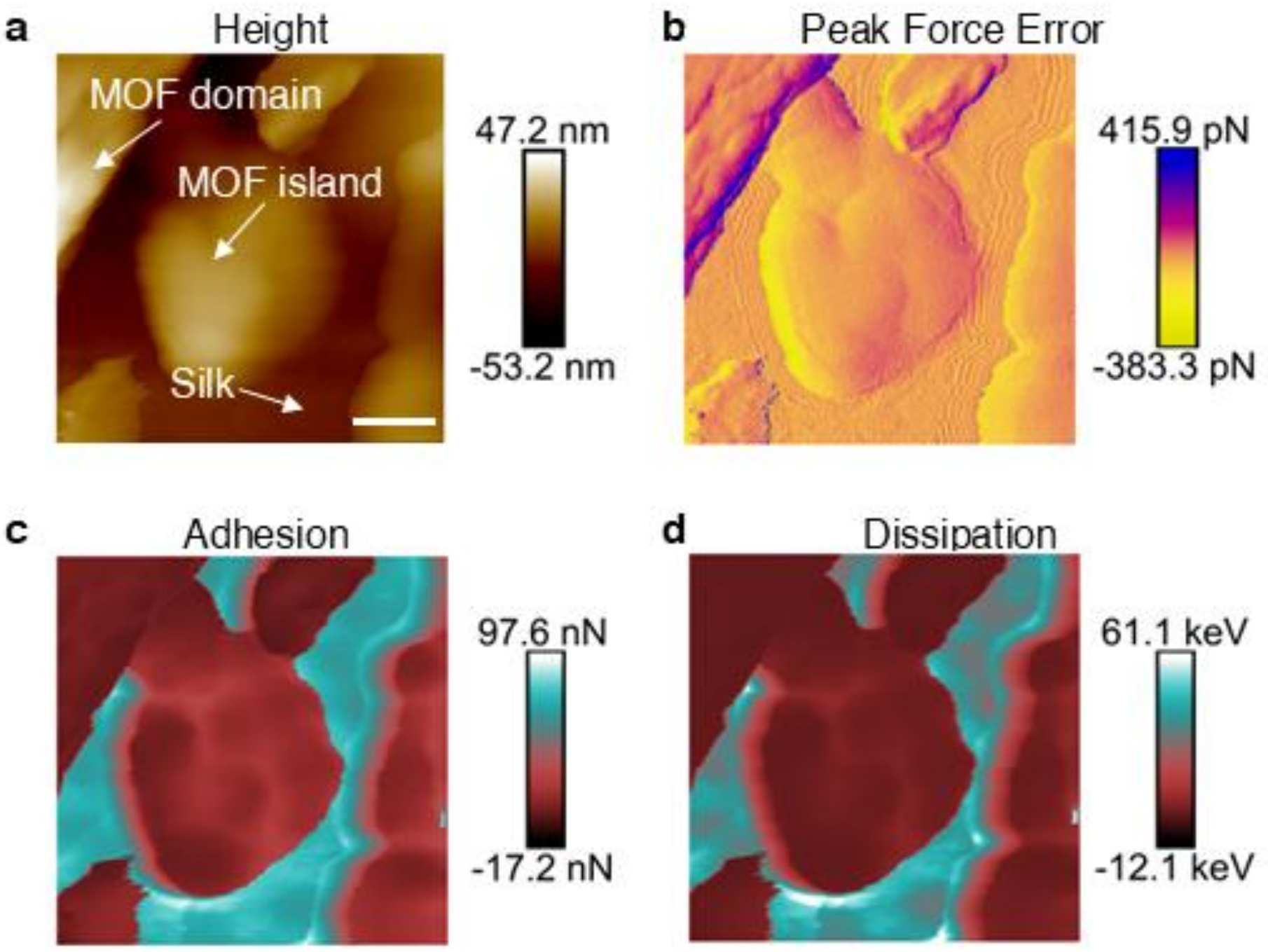

**Supplementary Figures 3 AFM PeakForce maps.** (a) Height. (b) Peak force error. (c) Adhesion. (d) Dissipation. Scala bar: 100 nm. The AFM maps show distinct features of nanomechanics between the MOF coating and the silk surface, which was applied to identify the boundaries between two materials.

Energy-dispersive X-ray spectroscopy (EDS) mapping was performed using SEM (GeminiSEM 300, Carl Zeiss, Germany) to examine the elemental composition of the coated fibres (Supplementary Fig. 4). Owing to the limited spatial resolution of EDS at this scale, the nanostrip pattern itself could not be directly resolved. Nevertheless, the elemental maps confirm the presence of Zr and N associated with the UiO-66-NH2 coating and support the overall distribution of the MOF phase along the fibre surface at the yarn scale.

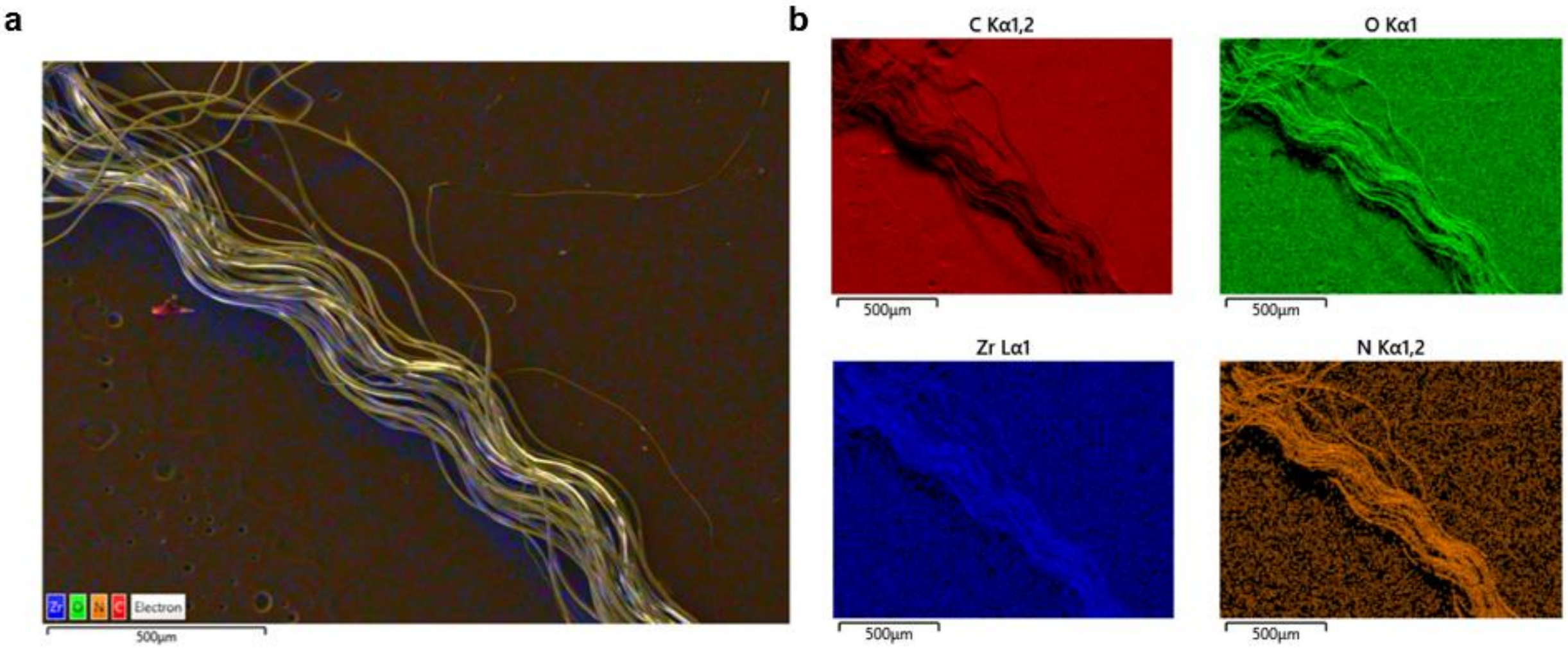

**Supplementary Figures 4 EDS maps of MOF nanosrip-coated silk fibres.** (a) Composite elemental map displaying overlaying the distributions of C, O, Zr, and N. (b) Individual elemental maps for C, O, Zr, and N, indicating the incorporation of the metal cluster of UiO-66-NH2 along the fibre surface.

To further analyse the secondary structure of the silk substrate, Fourier-transform infrared (FTIR) spectra of the silk fibre were deconvoluted in the amide I region (Supplementary Fig. 5). The fitted component peaks were assigned to β-sheet, random coil, α-helix, β-turn and side-chain contributions. Analysis of the integrated peak areas indicated a high crystalline fraction, consistent with the large β-sheet content of the silk template. This high intrinsic crystallinity is relevant to the selective surface chemistry and structural stability of the coated fibres.

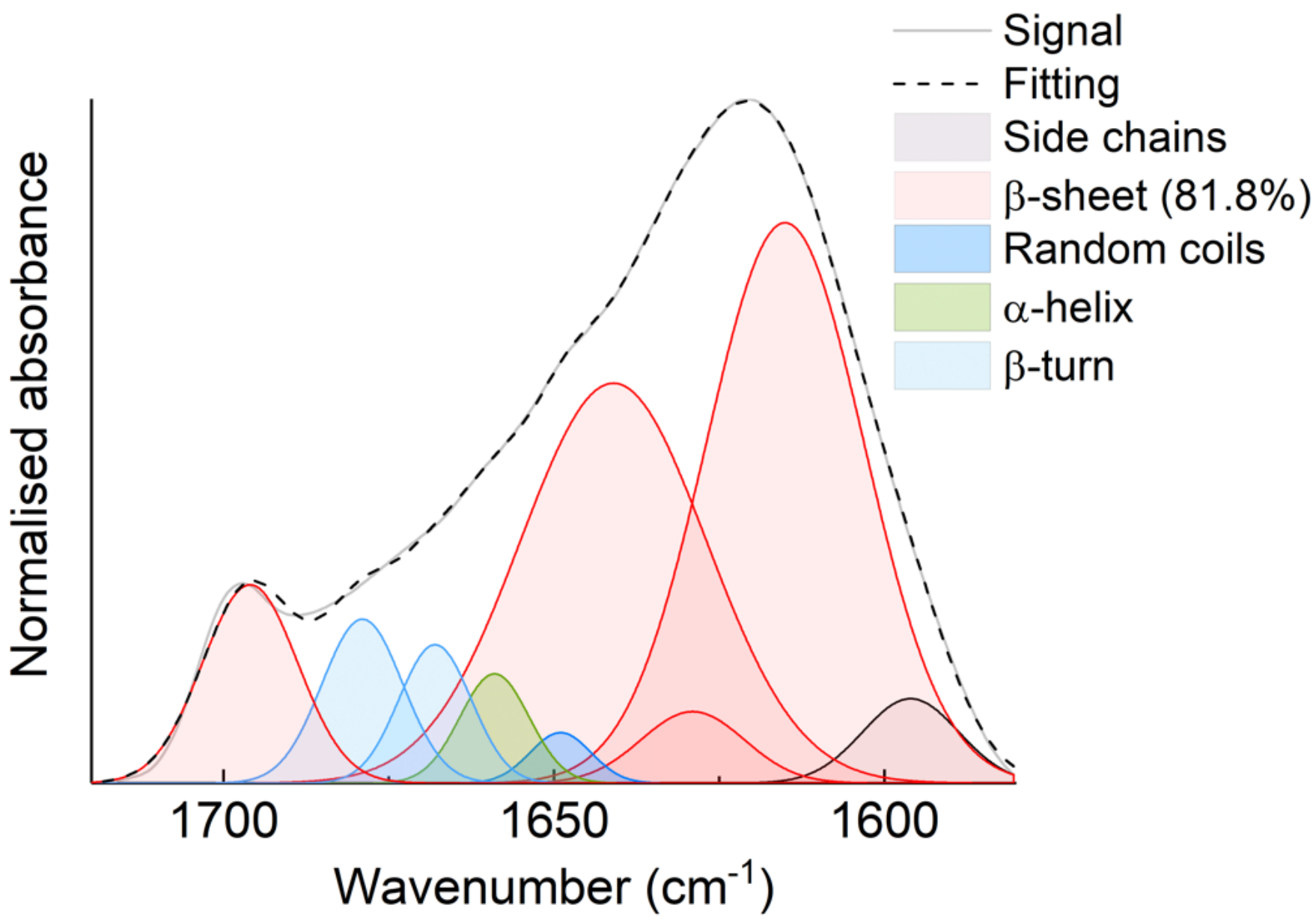

**Supplementary Figures 5 FTIR spectra of silk fibre and its deconvolution peaks.** The deconvolution peaks are centered at 1596 $cm^{-1}$ (side chains), 1615 $cm^{-1}$ (β-sheet), 1629 $cm^{-1}$ (β-sheet), 1641 $cm^{-1}$ (β-sheet), 1649 $cm^{-1}$ (random coils), 1659 $cm^{-1}$ (α-helix), 1668 $cm^{-1}$ (β-turn), 1679 $cm^{-1}$ (β-turn), and 1696 $cm^{-1}$ (β-sheet—intermolecular) [4,5]. A crystalline fraction as high as 81.8% was determined by analysing the areas of the deconvolution peaks in the Amide I region.

Repeated mechanical testing was carried out using a universal testing machine (Instron 5969) to assess the reproducibility of the tensile response of the fibres (Supplementary Fig. 6). These measurements complement the representative tensile behaviour shown in the main text and demonstrate consistent fracture behaviour across repeated tests.

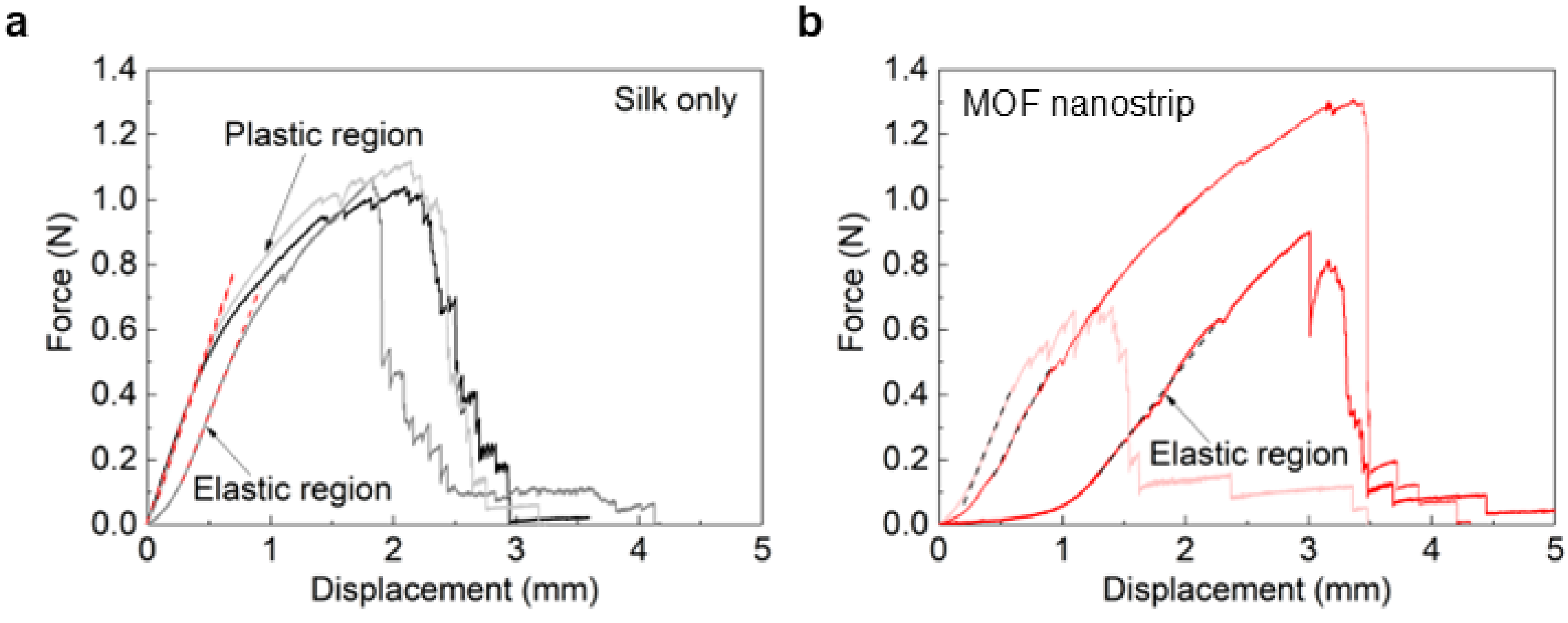

**Supplementary Figures 6 Repeated mechanical testing on fibres.** Dashed lines indicate the linear elastic region before the first break point.

Derivative thermogravimetric (DTG) curves resolve the mass-loss temperature ranges for the different samples (Supplementary Fig. 7). All samples display a ~60 °C peak attributed to water desorption. An additional ~210 °C peak in the MOF-only and MOF nanostrip-coated silk samples corresponds to MOF dehydration. All silk-containing samples exhibit a prominent peak at 310–340 °C, driven by silk substrate decomposition. At higher temperatures, the MOF-only sample showed a distinct peak near ~450 °C, whereas silk showed a peak near ~520 °C. The MOF nanostrip-coated silk exhibited an intermediate high-temperature peak near ~500 °C, suggesting coupled thermal decomposition behaviour arising from the coexistence of the thin MOF coating and the silk substrate.

DTG peak intensities further elucidate the interfacial structures. The ~60 °C water desorption peak intensity is highest for pure MOF, followed in descending order by pure silk, and the MOF nanostrip-, nanofilm-, and nanopowder-coated silks. This confirms the nanostrip sample possesses a thinner, more accessible MOF interface than the thicker controls. Similarly, the 310–340 °C peak intensity decreases progressively from bare silk to the MOF nanostrip-, nanofilm-, and nanopowder-coated samples, demonstrating that the nanostrip coating best preserves the intrinsic thermal behaviour of the underlying silk substrate.

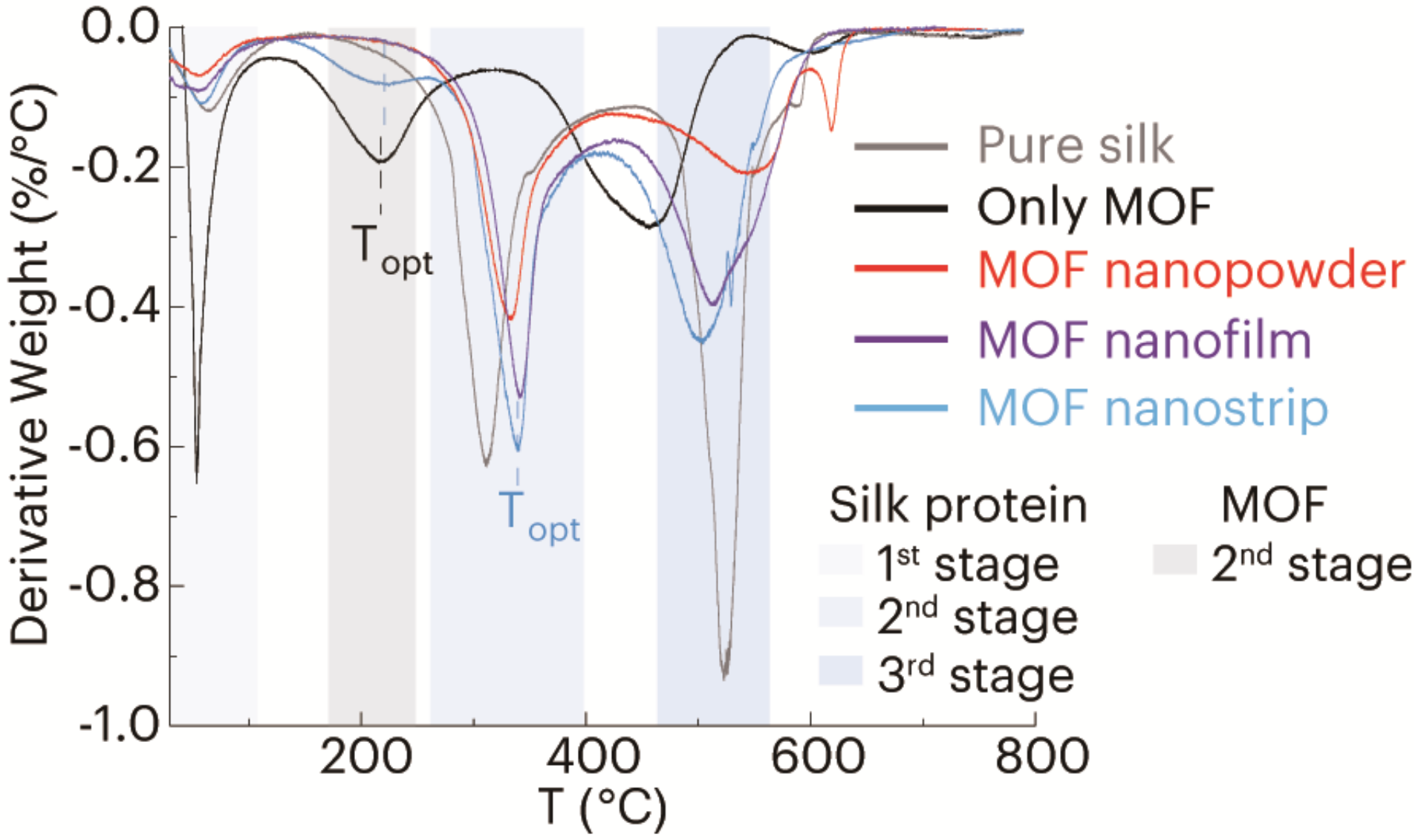

**Supplementary Figures 7 Derivative thermogravimetric curves**, showing the percentage of water mass loss ($m_{loss}$) at first stage and optimum degradation temperature ($T_{opt}$).

## Supplementary Note 3. Response time measurement

For response time measurement inside the environmental chamber, we designed a setup, including a controlled humid air flow and a PLA-printed chopper to periodically interrupt the flow, with frequency ranging from 1 to 50 Hz (See Supplementary Fig. 8). A pseudo four-point probe configuration was employed to minimise high-frequency inductive effects from the probes and measurement setup [6].

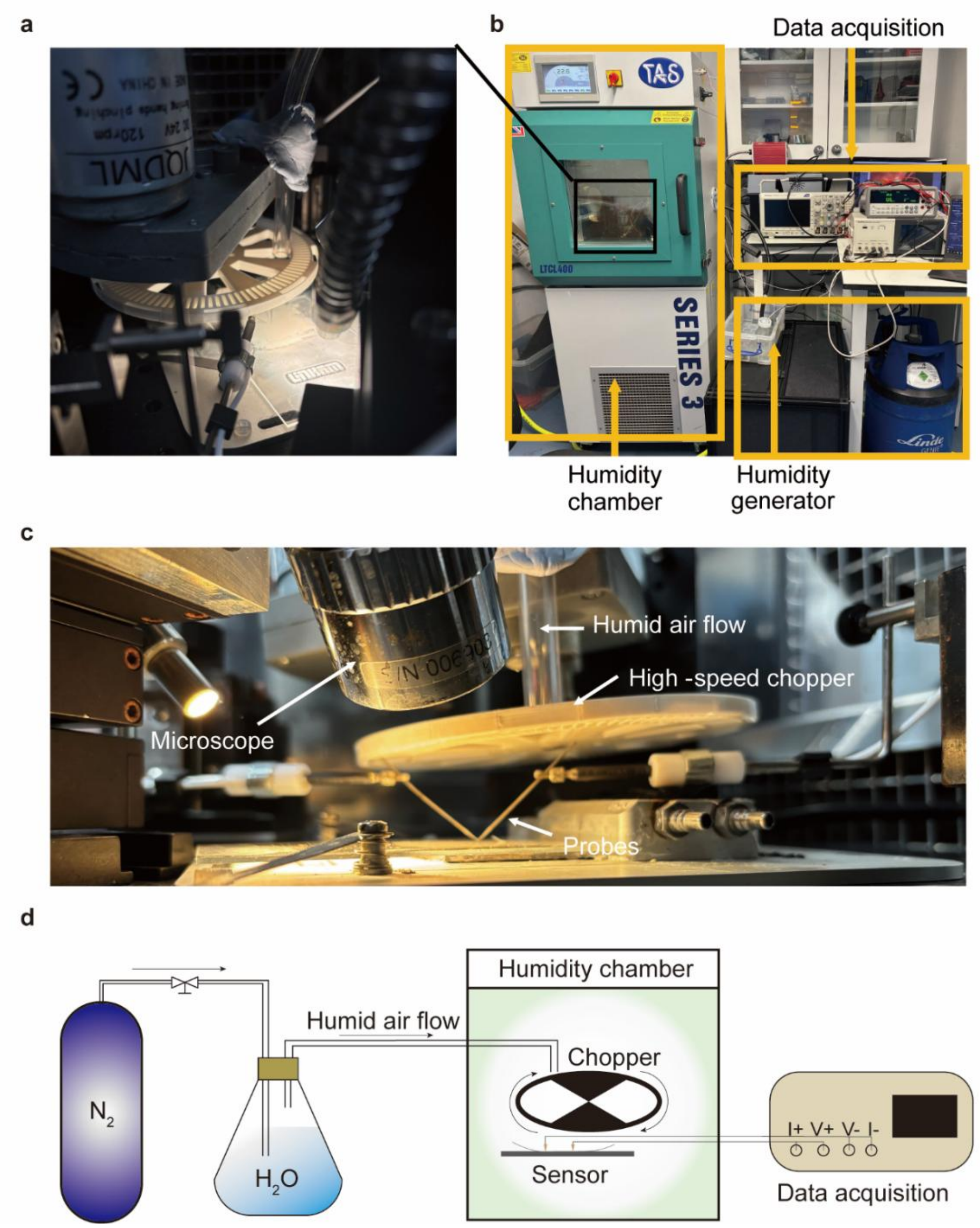

**Supplementary Figures 8 Humidity chamber testing setup.** (a) Photograph showing setup inside humidity chamber. (b) Testing system including humidity chamber, humidity generator and data acquisition devices. (c) Side-view of high-speed sensing measurement system. (d) Schematic of the setup for measuring sensor response speed.

## Supplementary Note 4. Electrochemical impedance spectroscopy analysis

### Electrochemical impedance fitting and equivalent-circuit analysis

Electrochemical impedance spectroscopy (EIS) data were analysed using IviumSoft (Ivium Technologies) to extract equivalent circuit parameters characterising the proton transport behaviour of different MOF-coated silk fibres. The impedance spectra were fitted using a Levenberg-Marquardt algorithm provided by IviumSoft. All fittings showed chi-square ($\chi^2$) values < 0.01, confirming the reliability and accuracy of the extracted circuit parameters.

A modified Randles circuit was used for MOF nanofilm and MOF nanopowder [7] (see Supplementary Fig. 9a), consisting of an ionic resistance ($R_s$) in series with a double-layer capacitance ($C_{dl}$), the interfacial charge transfer resistance ($R_{ct}$), and a Warburg impedance element ($Z_w$) accounting for ion diffusion, which is placed in series with $R_{ct}$. For MOF nanostrip, the equivalent circuit additionally included inductive components ($L_{ns}$) in parallel with a corresponding resistance ($R_{ns}$), due to the multi-threads features.

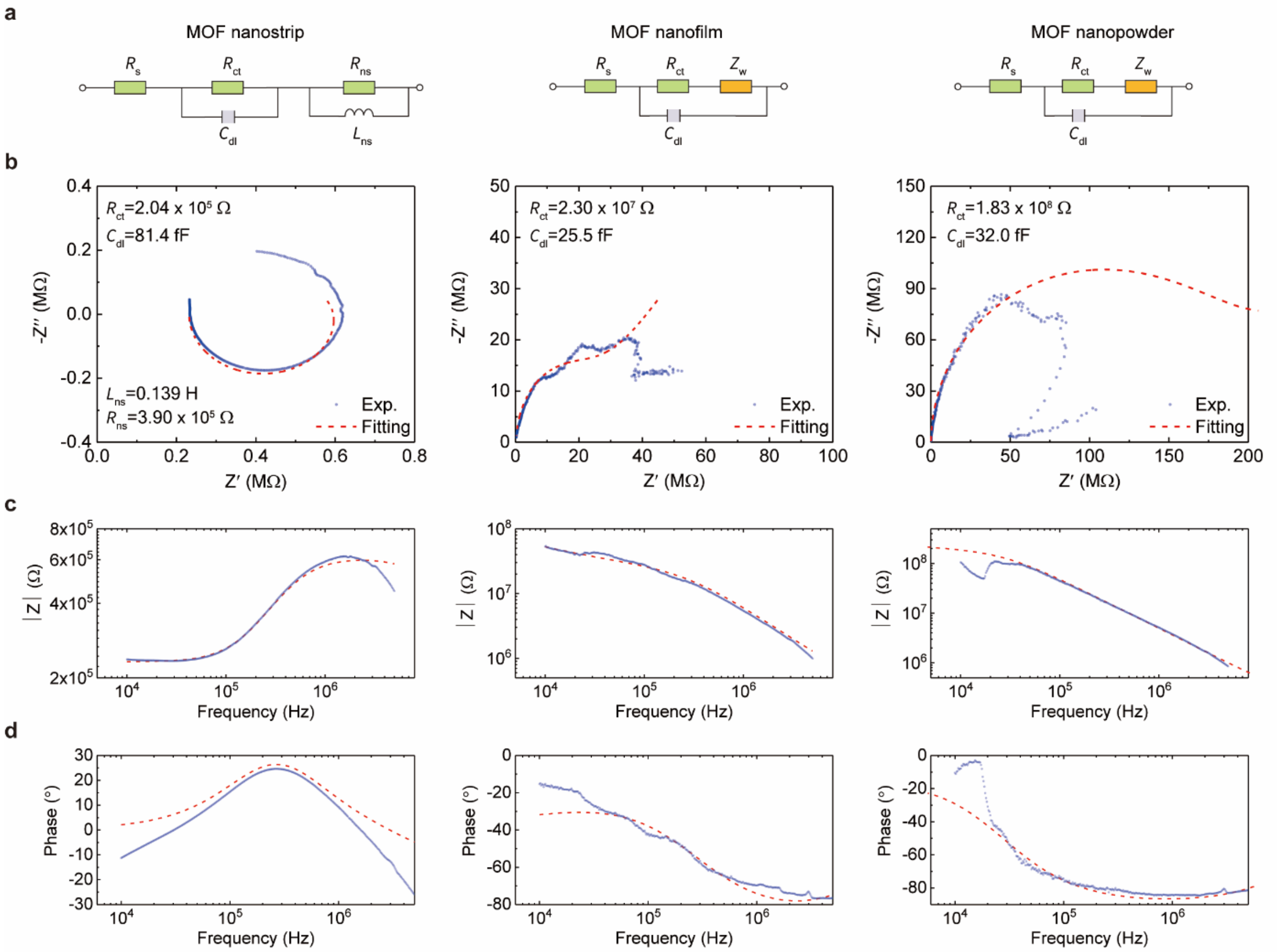

**Supplementary Figures 9 EIS equivalent circuit analysis**. (a) Schematic diagrams of the equivalent circuits used to model different sensor samples. (b) Nyquist plots of the measured

impedance spectra with fitted curves (red dashed lines), along with the extracted circuit parameters for each sample. (c, d) Corresponding Bode plots showing the impedance magnitude (c) and phase angle (d) as a function of frequency. MOF nanostrips exhibited a positive phase angle in the high-frequency region (> $10^5$ Hz), indicating a rea inductive behaviour rather than pseudo-inductance arising from interfacial polarisation delay [8].

## Humidity-dependent equivalent-circuit parameters

Supplementary Fig. 9 shows representative Nyquist plots and the corresponding fitted circuits for the three sensor architectures. Among them, the MOF nanostrip-coated sample exhibited the lowest $R_{ct}$, consistent with more efficient proton-conduction pathways enabled by the nanostrip architecture. The impedance spectra of this sample also displayed a distinct high-frequency inductive loop, which was absent in the MOF nanofilm and MOF nanopowder controls.

To evaluate the humidity dependence of the nanostrip-coated sensor, equivalent-circuit parameters were extracted for samples assembled with three growth cycles over a range of relative humidity values (Supplementary Fig. 10). The fitted parameters $(R_s, R_{ct}, R_{ns}, L_{ns}$ and $C_{dl})$ varied systematically with humidity, consistent with increased water uptake and enhanced proton mobility at higher RH. In particular, $\log_{10}(R_{ct})$ decreased approximately linearly with increasing RH, whereas $C_{dl}$ remained nearly constant at around $10^{-13}$ F. These trends support a humidity-activated proton-transport process dominated by reduced interfacial resistance rather than large changes in double-layer capacitance.

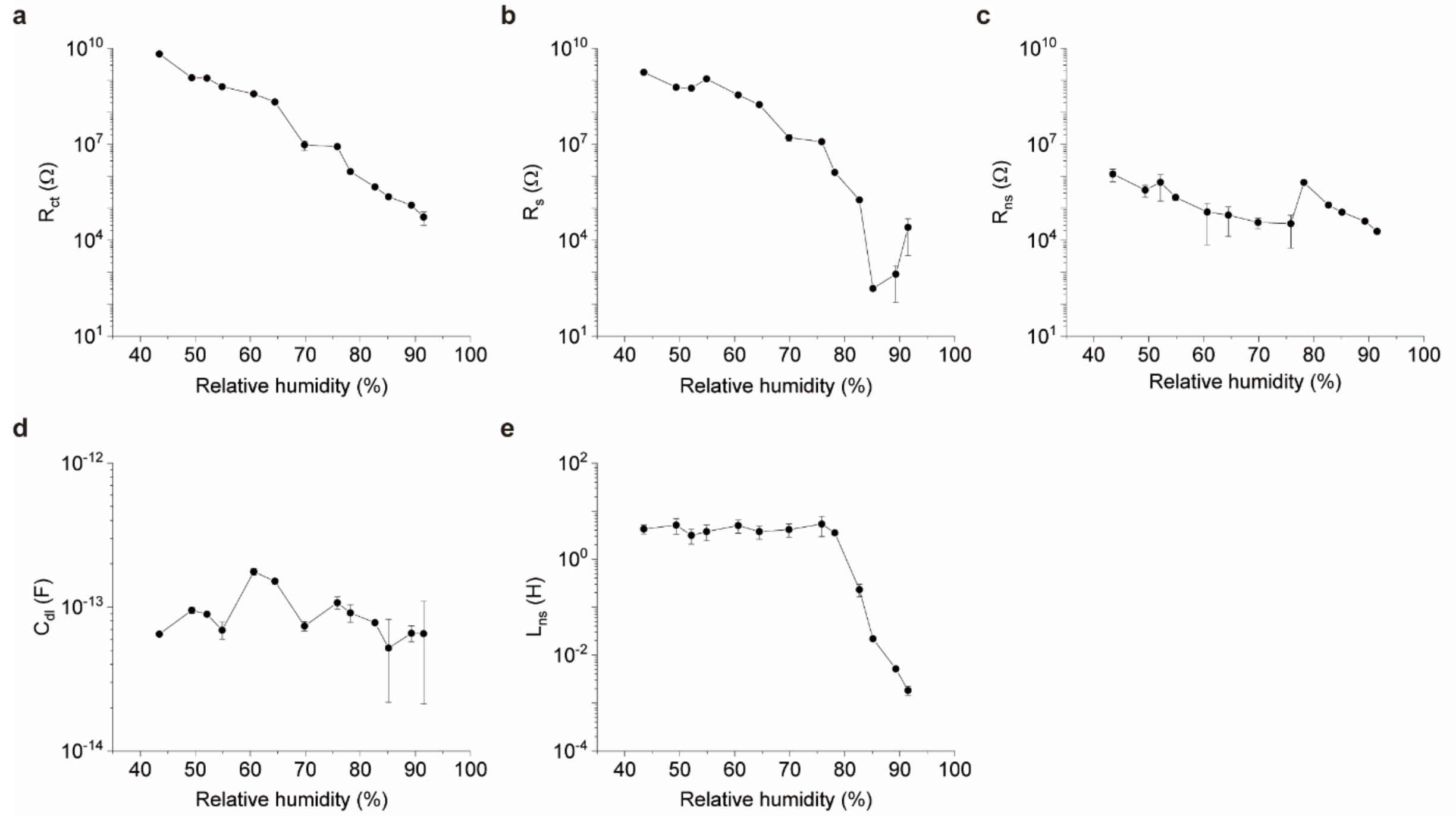

**Supplementary Figures 10 Humidity-dependent evolution of equivalent circuit parameters for MOF nanostrip-coated fibre**. Extracted values of (a) charge transfer resistance ($R_{ct}$), (b) series ionic resistance ($R_s$), (c) nanostrips-associated resistance ($R_{ns}$), (d) double-layer capacitance ($C_{dl}$), and (e) inductance component ($L_{ns}$) as a function of relative humidity. Data were obtained from EIS fitting of samples under varying humidity conditions.

## Nanostrip-specific inductive feature and characteristic relaxation time

A notable feature of the MOF nanostrip-coated samples was the presence of an equivalent inductive term $(\mathrm{L_{ns}})$, which was not required to fit the nanofilm or nanopowder controls. This observation indicates that the inductive feature is specific to the nanostrip structure rather than a general characteristic of MOF-coated fibre [7]. The fitted $\mathrm{L_{ns}}$ values were largest at low relative humidity and decreased strongly with increasing RH, with an approximately exponential drop above ~75% RH. This behaviour suggests that the nanostrip structure introduces a distinct interfacial relaxation process that is most pronounced under partially hydrated conditions and becomes less important once a more continuous proton-conduction network is established at higher humidity [9].

To further assess the timescale of this process, a characteristic relaxation time [10] was estimated as $\tau = \mathrm{L_{ns}}/\mathrm{R_{ns}}$. Using $\mathrm{R_{ns}}$ values on the order of $10^5$, the fitted $\mathrm{L_{ns}}$ values correspond to relaxation times ranging approximately from 10 μs to 10 ns, depending on humidity. These times indicate that the nanostrip-associated interfacial electrical relaxation is intrinsically fast and is consistent with the experimentally observed rapid sensor response.

## Supplementary Note 5. Apparent proton conductivity and Arrhenius analysis

### Apparent proton conductivity calculations

The apparent proton conductivity of the MOF nanostrip-coated fibres was estimated from the measured resistance using

$$\sigma = \frac{1}{R}\frac{L}{A} \quad (6)$$

where $L$ is the probe spacing, $R$ is the measured resistance, and $A$ is the effective cross-sectional area for axial proton transport. Because the MOF formed a thin surface coating on a fibre-like substrate, a thin-shell approximation was adopted, with the coating thickness $t$ (~70 nm) much smaller than the fibre diameter $D$ (12.5 μm). Under the upper-bound assumption of complete circumferential coverage, the effective conducting area was approximated as

$$A \approx \pi D t \quad (7)$$

which gives

$$\sigma = \frac{1}{R}\frac{L}{\pi D t} \quad (8)$$

In this expression, $t$ corresponds to the average thickness of the MOF nanostrip coating determined from AFM, and $D$ corresponds to the fibre diameter. All apparent proton conductivity values reported in Extended Data Fig. 3d were obtained using this geometric approximation.

### Effect of partial nanostrip coverage

The complete-coverage approximation provides a convenient upper-bound estimate of the conducting area, but it does not fully capture the actual morphology of the nanostrip coating. AFM maps and SEM images indicate that, after three layer-by-layer assembly cycles, the MOF does not form a perfectly uniform shell. Instead, continuous nanostrips preferentially form on nanofibrillar regions, whereas the intervening amorphous regions contain only sparse and disconnected MOF islands. The effective circumference participating in axial conduction is therefore expected to be smaller than $\pi D$.

To account for this effect, a coverage factor $f (0 < f \leq 1)$ may be introduced and the corresponding conductivity becomes

$$\sigma = \frac{1}{R}\frac{L}{\pi D t f} \quad (9)$$

The value of $f$ can be estimated as 0.55, determined from the present data, and the true proton conductivity of the nanostrip regions is 1.8 times higher than the apparent values reported here, while remaining within the same overall order of magnitude.

### Effect of interfacial water layers at high humidity

At high relative humidity, axial proton transport may also involve a hydrated interfacial layer associated with adsorbed water. Literature on hydrophilic interfaces generally indicates that such adsorbed water layers are typically molecularly thin, often on the order of ~1–3 nm under humid conditions [11-13], and therefore remain substantially thinner than the ~70 nm MOF nanostrip thickness. Accordingly, the contribution of an interfacial water layer is expected to modify, but not dominate, the effective transport cross-section. If an additional hydrated layer of thickness $t_w$ is included, the corresponding conductivity becomes

$$\sigma = \frac{1}{R}\frac{L}{\pi D(t+t_w)} \tag{10}$$

For representative values of $t_w$ = 1-3 nm and $t \approx 70$ nm, the inclusion of the hydrated interfacial layer would increase the assumed conducting thickness by only ~1.4–4.3%, corresponding to a similar decrease in the calculated conductivity. Thus, although the measured value should still be regarded as an apparent proton conductivity under humid operating conditions, the correction associated with a nanometre-scale interfacial water layer is modest and does not alter the overall order of magnitude.

This situation differs from that of conventional bulk conductivity measurements on compressed powder pellets [14] or large microcrystals [15], in which the geometric cross-section is dominated by the macroscopic sample dimensions and any surface water layer makes a negligible relative contribution. In the present surface-confined nanostrip architecture, the same water layer is more relevant in relative terms, but it remains small compared with the nominal coating thickness and therefore does not qualitatively change the conclusion of exceptionally high proton transport.

### Comparison with literature

Supplementary Table 1 compares our results with literature values for conventional proton-conducting materials, which are typically evaluated in bulk geometries (e.g., pellets, single crystals, or thick films). Unlike these bulk systems, our material relies on an ultrathin, surface-confined nanostrip coating. Consequently, our measurements represent apparent proton conductivities derived from nominal coating thickness. Despite this geometric distinction, the nanostrip conductivities substantially exceed those of conventional MOF formats. This confirms that the longitudinal, highly accessible nanostrip architecture enables exceptionally efficient proton transport.

**Supplementary Table 1 Comparison of representative proton conductivities reported in the literature**

| Material | Proton conductivity (S cm$^{-1}$) | Temperature (°C) | RH (%) | Ref. |
|---|---|---|---|---|
| Protonic β-aluminas | $10^{-3}$–$10^{-4}$ | 300 | – | [16] |
| Anatase | $3.78 \times 10^{-2}$ | 80 | 81 | [17] |
| Nafion/α-ZrP | 0.1 | 100 | 100 | [18] |
| GO | $1.0 \times 10^{-2}$ | 26.85 | 100 | [19] |
| GO–Nafion | 0.047 | 120 | 30 | [20] |
| Imidazole/Nafion | 0.1 | 160 | 100 | [21] |
| sPEEK/sDPSD | 0.1 | 120 | – | [22] |
| $(NH4)_2(adp)[Zn_2(ox)_3]\cdot 3H_2O$ (MOF) | 0.8 | 25 | 98 | [23] |
| ZST MOFs | $2.4 \times 10^{-3}$–$5.6 \times 10^{-3}$ | 65 | 95 | [24] |
| MOF (Cu–TCPP) | $3.9 \times 10^{-3}$ | 20 | 98 | [25] |
| $H_2SO_4$@MIL-101 | $1.0 \times 10^{-2}$ | 150 | 20 | [26] |
| $[Al(\mu_2\text{-}OH)(1,4\text{-}ndc)]_n$ | $2.2 \times 10^{-5}$ | 120 | – | [27] |
| $Fe(ox)\cdot 2H_2O$ | $1.3 \times 10^{-2}$ | 25 | 98 | [28] |
| Trz@TPB-DMTP-COF | $1.1 \times 10^{-3}$ | 130 | – | [29] |
| CNT/NKCOFs | $1.0 \times 10^{-3}$ | 24.85 | – | [30] |
| MOF nanostrip-coated silk fibre | 39.7 (UiO-66-OH) | 20 | 83 | This work |

## Activation-energy analysis

Resistance measurements on nanostrip-coated silk were performed at 1, 20, 40, 60 and 80 °C in humidity chamber. At each temperature, the proton conductivity was calculated according to Eq. 8. The activation energy, $E_a$, was determined using the Arrhenius relationship

$$\sigma = \sigma_0 exp\left(-\frac{E_a}{k_B T}\right), \tag{11}$$

where $\sigma_0$ is the pre-exponential factor, $k_B$ is the Boltzmann constant and T is the absolute temperature. Accordingly,

$$\ln \sigma = \ln \sigma_0 - \frac{E_a}{k_B T} \tag{12}$$

$E_a$ was obtained from the slope of the linear fit of $\ln \sigma$ against 1000/T (shown in Figure 3g).

## Supplementary Note 6. Robustness and stability tests

To assess mechanical and environmental stability, MOF nanostrip-coated sensing fibres were subjected to various robustness tests (see Supplementary Fig. 11 and Supplementary movie 1). Impact resistance was tested using a custom nitrogen-driven jet setup (2.5 mm nozzle, up to 10 bar, ~30 $m \cdot s^{-1}$) to assess the structural stability of the coatings under strong water jet [3,31]. Chemical stability was evaluated by soaking samples in pure ethanol at room temperature for one week. Washing durability was assessed using a household rotary washing machine operated for 3 h. Mechanical flexibility was tested *via* 1,000-cycle bending using a universal testing machine (Instron). After each test, samples were vacuum-dried overnight at 100 °C, and conductivity changes were recorded to evaluate performance retention.

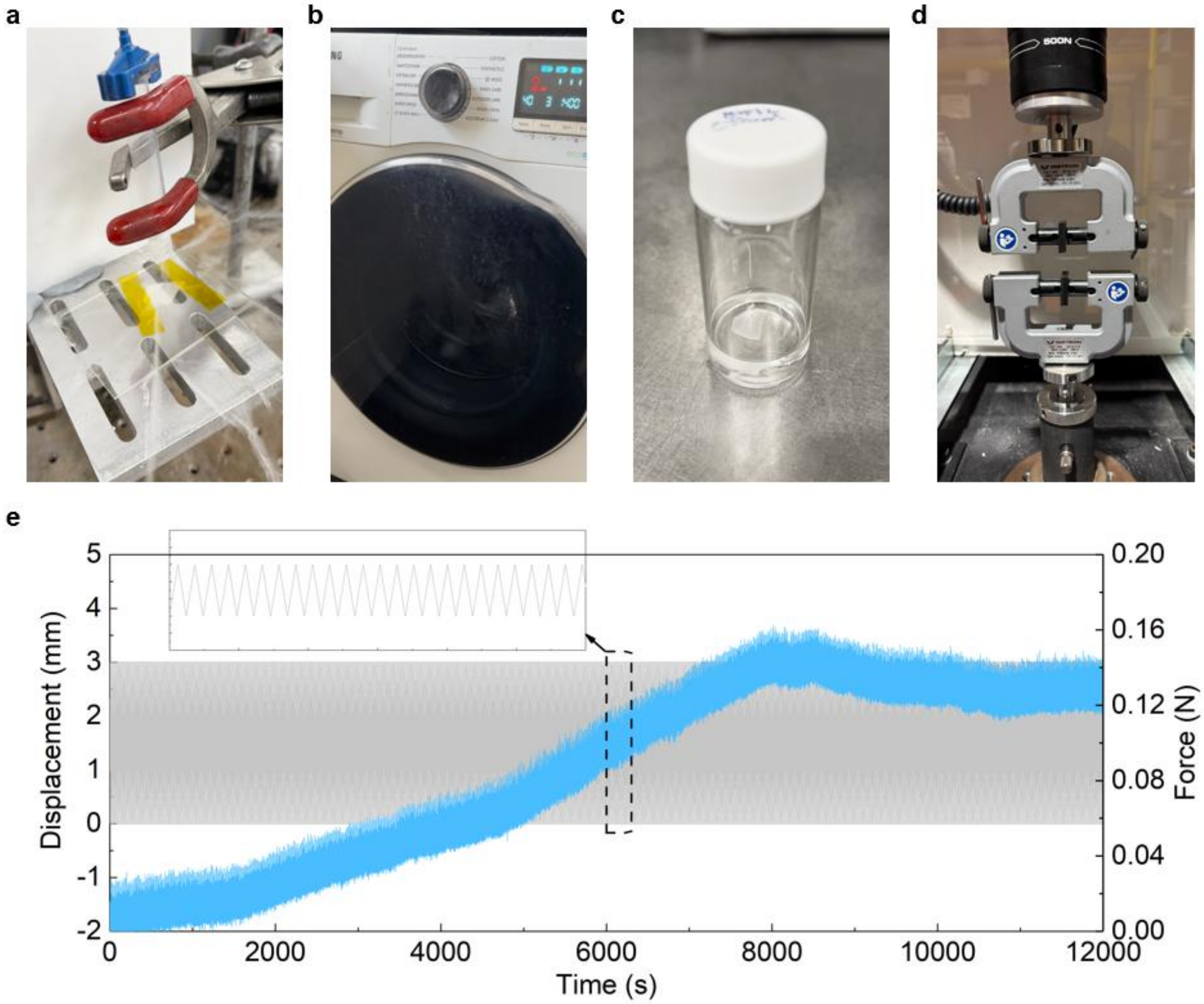

**Supplementary Figures 11 Robustness test on MOF nanostrip-coated silk textile and fibres.** (a) Photograph of high-speed water jet impact test. (b) Photograph of household washing machine tests. (c) Photograph of textile after immersion in ethanol solution for one week. (d) Photograph of cyclic bending tests conducted with a universal testing machine, accompanied by (e) the corresponding displacement (grey) and force (blue) curves.

## Supplementary Note 7. Cytotoxicity test

### Cell Culture and Treatment

Fabric samples (4 × 4 cm) were incubated in a Dulbecco's Modified Eagle Medium supplemented with 10% fetal bovine serum and 1% penicillin-streptomycin at 37 °C for 7 days to obtain the leach liquor. Pure silk fabrics were used as a control group. L929 fibroblast cells were seeded in a 48-well plate at a density of $3 \times 10^4$ cells per well and allowed to adhere overnight in a humidified incubator maintained at 37 °C with 5% $CO_2$. For treatment, 400 μL of the fabric leachate and 100 μL of the cell culture medium were added to the wells.

### Lactate Dehydrogenase (LDH) Cytotoxicity Assay

After 7-day incubation, 50 μL of the culture supernatant from each well was transferred to a 96-well plate, followed by the addition of 50 μL of LDH assay reagent. The plate was incubated at 37 °C for 30 mins in the dark. After the addition of stop solution, absorbance was measured at 490 nm using a microplate reader (Tecan Infinite Nano, Switzerland). Statistical analysis was conducted using GraphPad Prism software, and results were reported as mean ± standard deviation (SD) from three independent experiments.

### Live/Dead Staining and fluorescence Imaging

LIVE/DEAD assay working solution was prepared by mixing 5 μL of calcein-AM (Component A) and 20 μL of ethidium homodimer-1 (EthD-1, Component B) into 10 mL of phosphate-buffered saline (PBS, pH 7.4, Merck, USA) solution. After 7 days of treatment, cells were washed twice with PBS and incubated with the staining solution at 37 °C for 30 min in the dark. Excess dye was then removed by gentle PBS rinsing, and fresh culture medium was added prior to imaging. Fluorescence images (Supplementary Fig. 12) were acquired using an Aurox Clarity confocal fluorescence microscope (Aurox Ltd., UK).

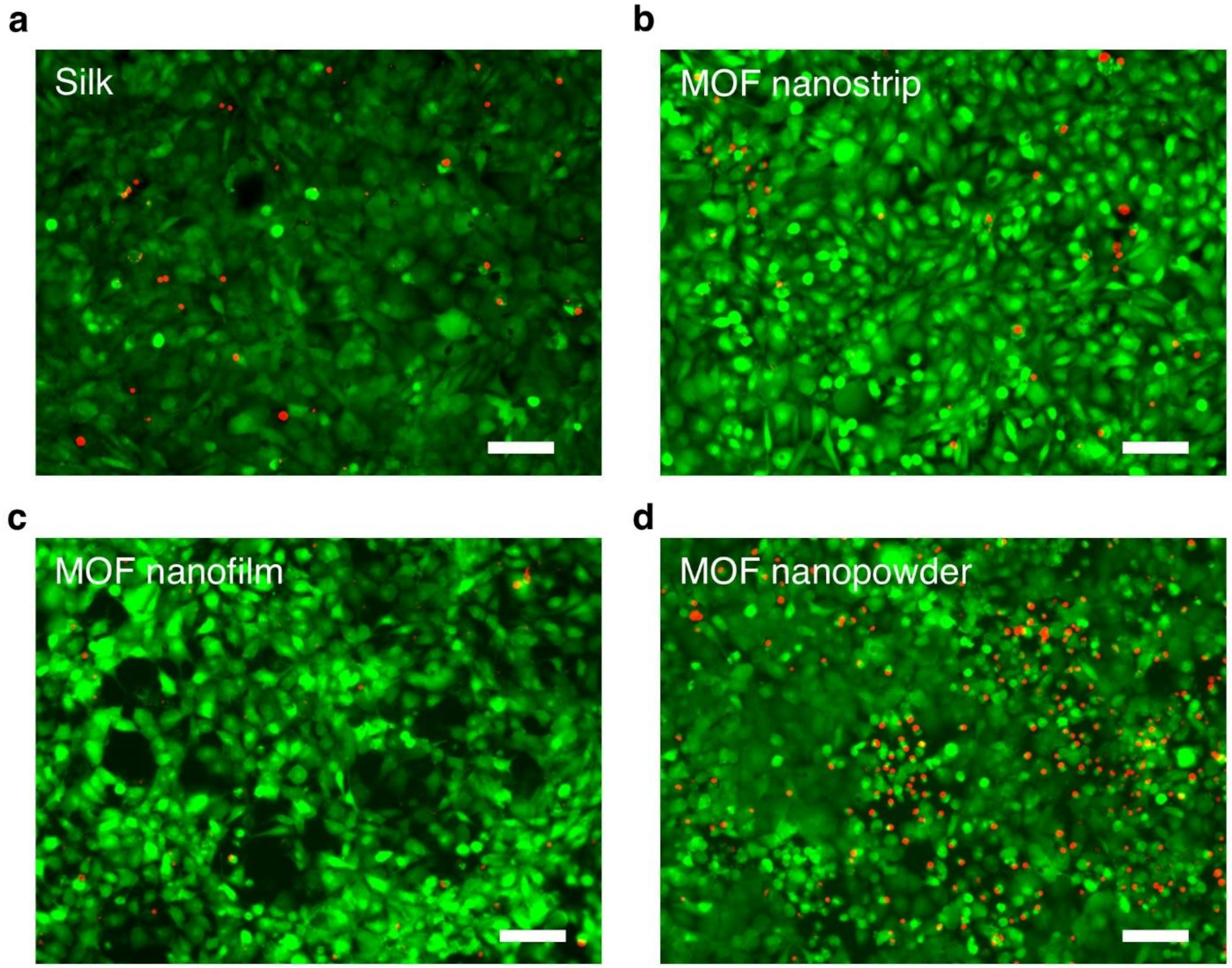

**Supplementary Figures 12 Fluorescence microscopy images for cytotoxicity test.** L929 fibroblast cells after 7 days of treatment with different extract of fabric materials. (a) Cultured cells treated with silk extract. (b) Cells treated with MOF nanostrip extract. (c) Cells treated with MOF nanofilm extract. (d) Cells treated with MOF nanopowder extract. Live cells are stained green and dead cells red. Scale bar: 50 μm.

## Supplementary Note 8. Electrode integration of the single-thread sensor

To convert the MOF-coated silk yarn into a complete respiratory sensing thread, electrical contact was established at both ends of the sensing element using two simple electrode-integration strategies: conductive silver ink painting and conductive wire knotting (Supplementary Fig. 13). In both cases, the objective was to provide low-resistance electrical connections while preserving the exposed MOF-coated sensing region for humidity detection.

For the first approach, conductive silver ink was applied directly to both ends of the MOF-coated silk yarn to form terminal electrodes (Supplementary Fig. 13a-d). After drying, the silver-coated regions served as conductive contact pads for external electrical connection. The resistance of the silver-ink electrode alone was ~33 Ω, whereas the resistance of the complete device under nasal breathing remained in the megaohm range (~5.2 MΩ), indicating that the electrode contribution was negligible compared with that of the active sensing region. For the second approach, conductive silver wires were mechanically connected to both ends of the sensing yarn by knotting (Supplementary Fig. 13e-h). This method avoided direct coating of the terminal region with conductive paste and provided a straightforward textile-compatible route for device assembly. The resistance of the knotted wire electrode alone was ~15.4 Ω, while the resistance of the whole device under nasal breathing was ~5.3 MΩ.

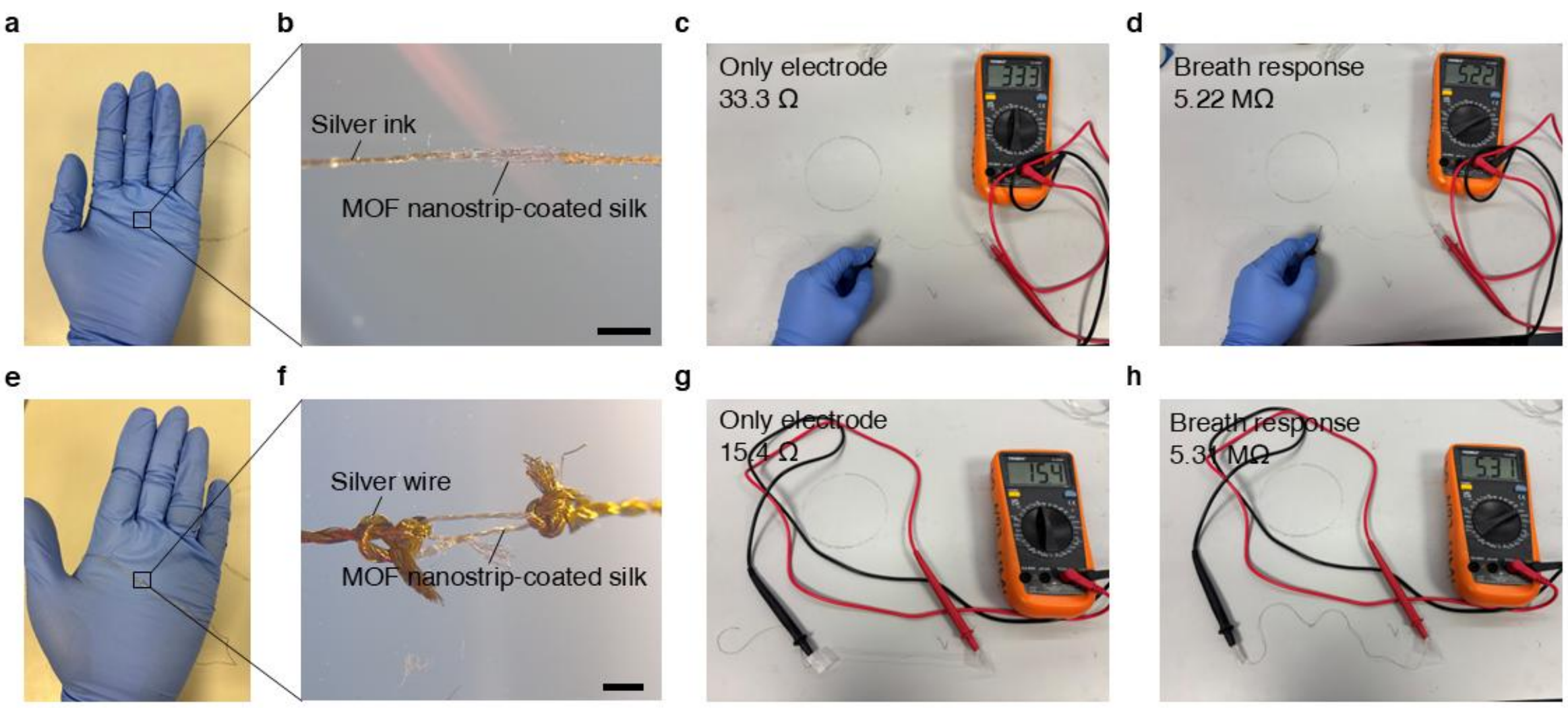

**Supplementary Figures 13 Electrode configurations on the single-thread respiratory sensor.** (a) Photograph and (b) optical microscopy image of a sensor with electrodes formed by painting conductive silver ink. Scale bar: 1 mm. (c) Resistance of the silver ink electrode alone (~ 33 Ω), and (d) resistance of the complete sensor (~ 5.2 MΩ) under nasal breathing. (e) Photograph and (f) optical microscopy image of a sensor with silver wires connected by knotting. Scale bar: 1 mm. (g) Resistance of the knotted silver wire electrode alone (~ 15.4 Ω), and (h) resistance of the whole sensor (~ 5.3 MΩ) under nasal breathing.

## Supplementary Note 9. Human volunteer testing protocol under ambient conditions

A human pilot study was conducted involving 17 healthy adult participants (demographics detailed in Supplementary Table 2) to evaluate the performance of the MOF nanostrip-coated silk humidity sensor for real-time respiratory monitoring. Experiments were performed in a climate-controlled laboratory maintained at 20°C and 40% RH. The single-thread MOF sensor was integrated into a standard nasal cannula and positioned at the nasal vestibule to monitor expiratory and inspiratory airflow. Participants were seated in a resting state and guided through a structured protocol of controlled breathing tasks: (1) Baseline Monitoring: Normal breathing (2 min) and slow/flow breathing (30 s); (2) Vocal Activity: Humming (1 min) to assess sensor response to turbulent airflow and frequency changes; (3) Long-term Stability: continuous monitoring during normal activity for 30 minutes.

To validate the sensor's temporal precision, participants synchronised their breathing with a programmed metronome (Supplementary Fig. 14) at preset respiratory rates. All data were recorded under professional supervision to ensure physiological safety and data integrity.

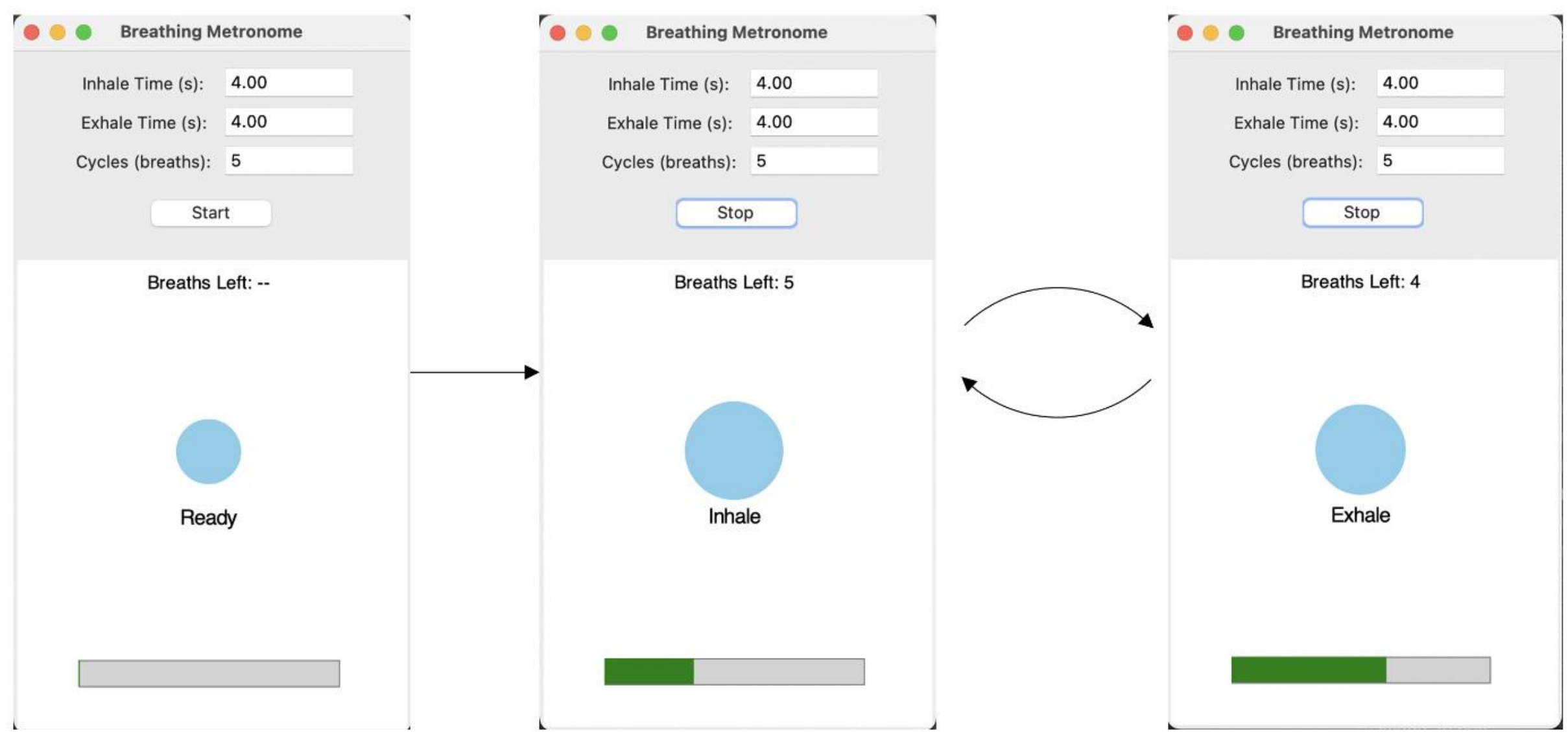

**Supplementary Figures 14 Metronome interface for volunteer test.** The user interface displays adjustable parameters including inhale time, exhale time, and number of breath cycles. A visual progress bar guides participants in synchronising their breathing with the predefined rhythm.

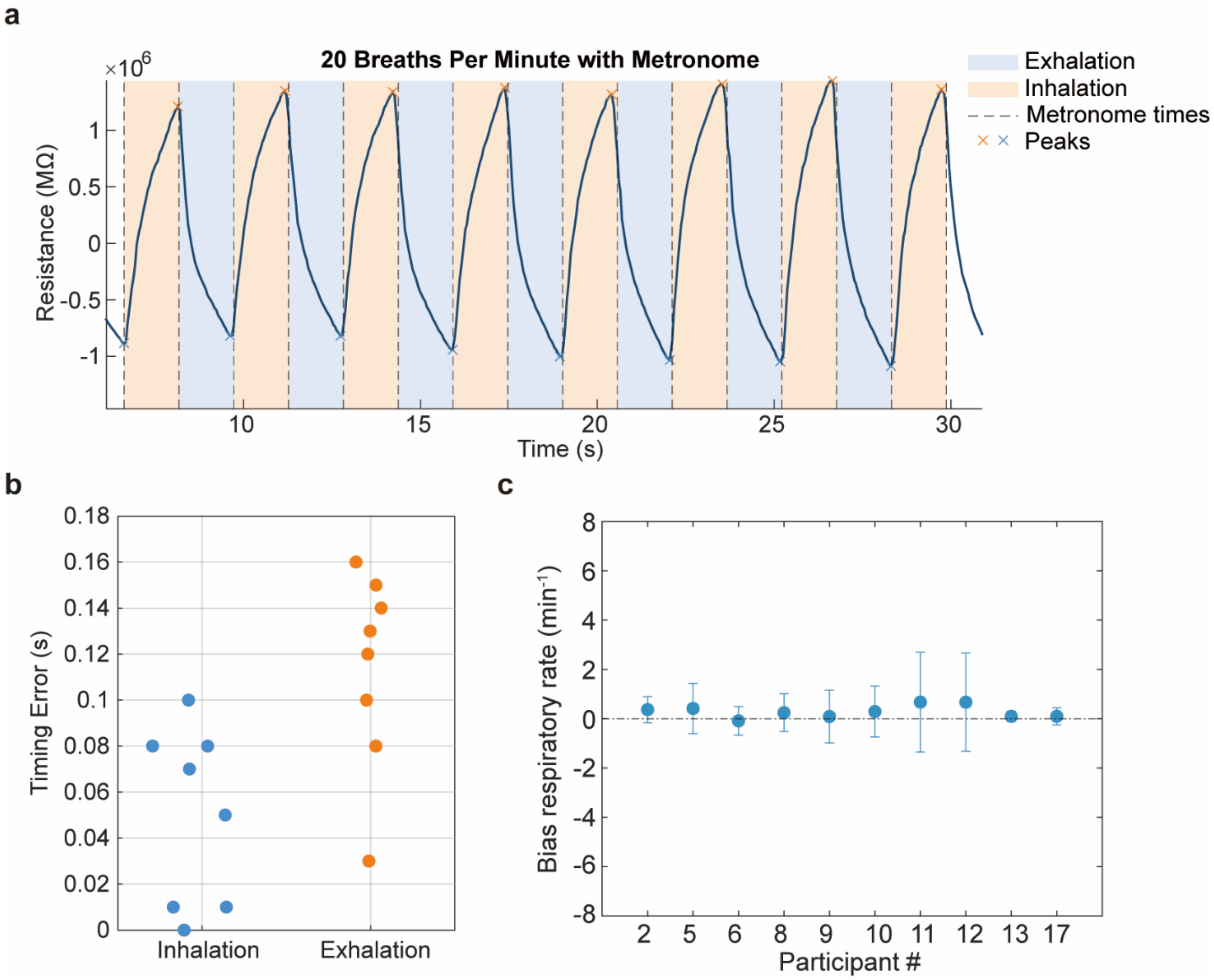

**Supplementary Figures 15 Breathing monitoring results by asking a participant following a metronome.** (a) Real-time resistance changes of the sensor as a volunteer breathes at a predefined respiratory rate synchronised with a metronome. The raw data were detrended to remove slow baseline drift, primarily arising from circuit instability, for clarity in periodic signal analysis. (b) Time difference between the peaks of the recorded breathing waveform and the corresponding metronome beats. (c) Deviation between the actual breathing rates measured by the sensor and the target rates set by the metronome across multiple participants.

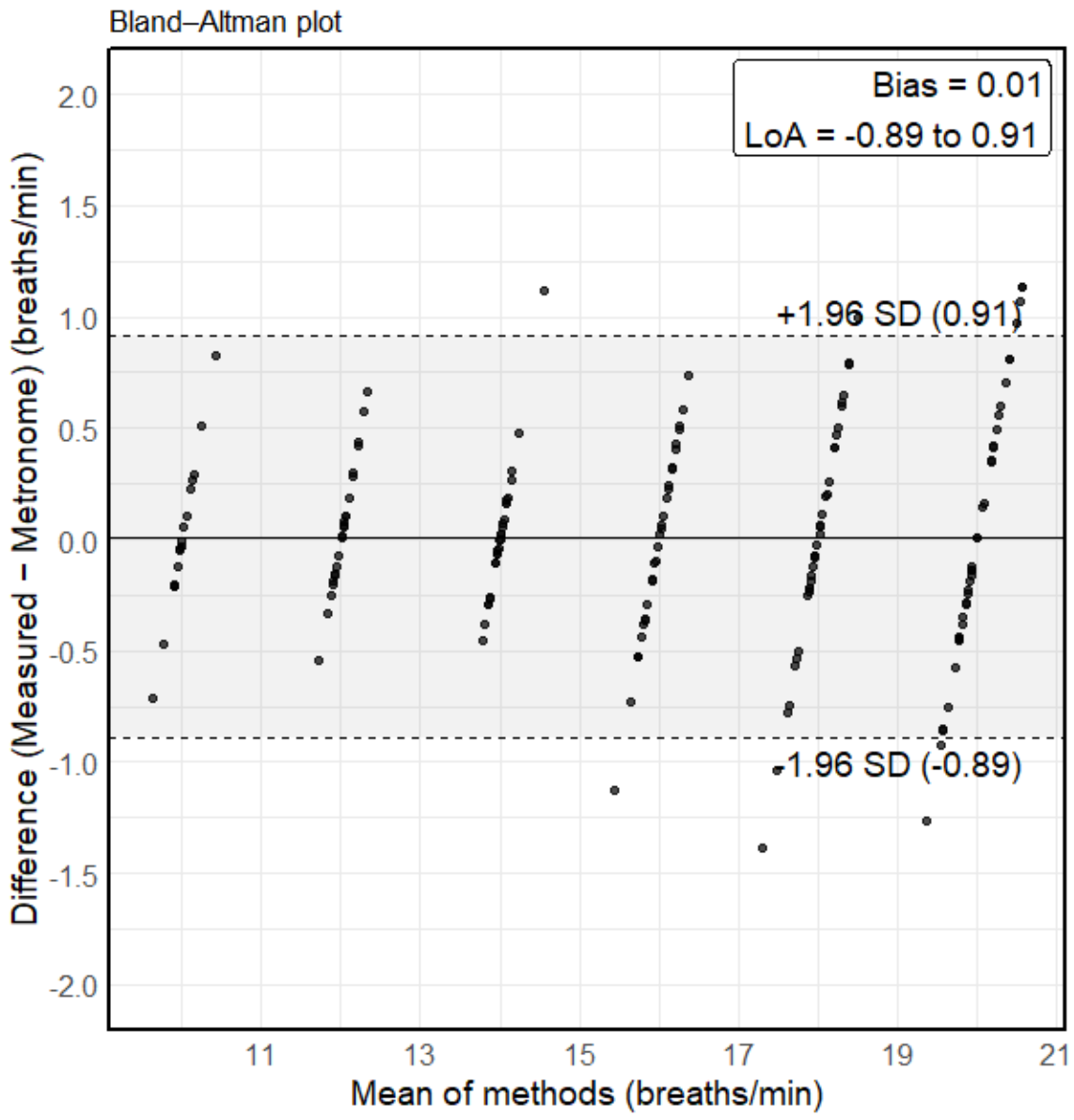

**Supplementary Figures 16 Bland–Altman plot** comparing measured respiratory rate with the metronome reference. Differences are plotted against the mean of the two methods. The solid line shows the bias (0.01 breaths/min; 95% CI −0.06 to 0.08); dashed lines mark the 95% limits of agreement (±1.96 SD: −0.89 to 0.91 breaths/min; 95% CIs: lower -1.011 to -0.773, upper 0.793 to 1.031). n = 171.

**Supplementary Table 2 Information about human volunteers**

| Participant number | Sex | Age | Symptoms |
|---|---|---|---|
| 1 | M | 27 | No |
| 2 | F | 27 | Viral coryzal symptoms |
| 3 | M | 25 | No |
| 4 | M | 29 | No |
| 5 | M | 33 | No |
| 6 | M | 37 | No |
| 7 | F | 44 | No |
| 8 | M | 35 | No |
| 9 | F | 50 | Cold |
| 10 | F | 30 | No |
| 11 | F | 30 | No |
| 12 | F | 28 | No |
| 13 | M | 62 | No |
| 14 | M | 29 | Cold |
| 15 | M | 30 | Nausea |
| 16 | M | 30 | No |
| 17 | M | 26 | No |

M: male; F: female.

**Supplementary Note 10. Wireless node and clinical interface for single-thread respiratory monitoring**

A miniaturised wireless node was fabricated to enable real-time respiratory monitoring using the MOF nanostrip-coated sensor. The silver conductive tracks of the circuit were printed on a polyimide (PI) substrate (Goodfellow) by a direct writing process. This was done using a fixed syringe loaded with a commercial silver ink (DM-SIP-2005) and the substrate loaded on the XYZ Aerotech nanopositioning system (ANT130L and ANT130XY). The stage translated beneath the fixed needle to allow for programmable deposition of conductive traces on the flexible substrate.

Passive surface-mount components, including resistors and capacitors, together with a low-power operational amplifier (MCP6001T-I/OT), were assembled directly onto the printed circuitry and bonded using the same silver ink. The completed analog front-end was connected to an ESP32C3 Bluetooth chip for wireless communication and powered by a Li-ion battery (RS Components, 190 mAh, model HPL402323-2C). The assembled device and corresponding circuit schematic are shown in Supplementary Fig. 17a–c.

The wireless node transmitted respiratory raw data to a computer *via* BLE, where a custom Python-based graphical interface, *RespiraFibre* monitor, was used for real-time visualisation of the waveform and extraction of respiratory rate (RR). To improve clinical usability, the interface additionally displayed the NEWS2 respiratory-rate parameter and risk level, allowing direct interpretation of respiratory status in a clinically familiar format (Supplementary Fig. 17d).

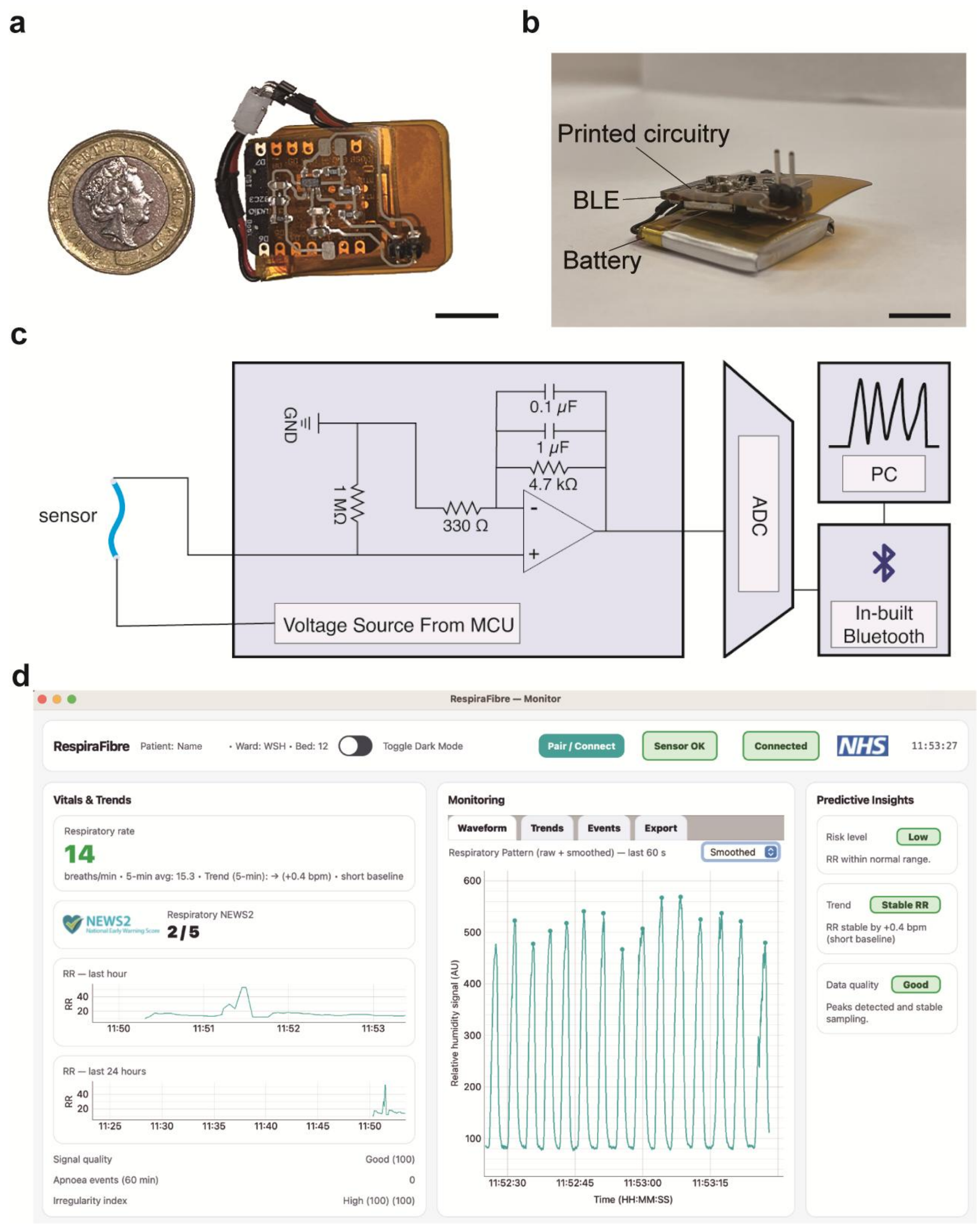

**Supplementary Figures 17 Wireless node for respiratory sensing.** (a) Photograph of the wireless node shown next to a one-pound coin for size comparison. Scale bar: 1 cm. (b) Side-view photograph of the wireless node, showing the stacked architecture of the printed circuitry, Bluetooth Low Energy (BLE) module and battery. Scale bar: 1 cm. (c) Circuit schematic of the wireless readout system, including the sensor, voltage source, amplifier, analog-to-digital converter (ADC) and integrated BLE communication module. (d) A custom Python-based clinical interface.

## Supplementary Note 11. High-flow oxygen therapy pre-clinical validation

To evaluate the sensor's resilience under extreme clinical conditions (i.e. high flow rates and near-saturated humidity), clinical validation was conducted at the University College London Hospital (UCLH) Intensive Care Unit (ICU). A healthy male volunteer (39 years old) participated in the high-flow oxygen therapy (HFOT) trials. During these sessions, a Maquet system for oxygen concentration and interface control, integrated with a Fisher & Paykel Healthcare humidification system (Optiflow) and high-flow breathing circuits were employed (see Figure 4g and Supplementary Fig. 18). This setup delivered opti-humidified oxygen at 37°C and 100% RH.

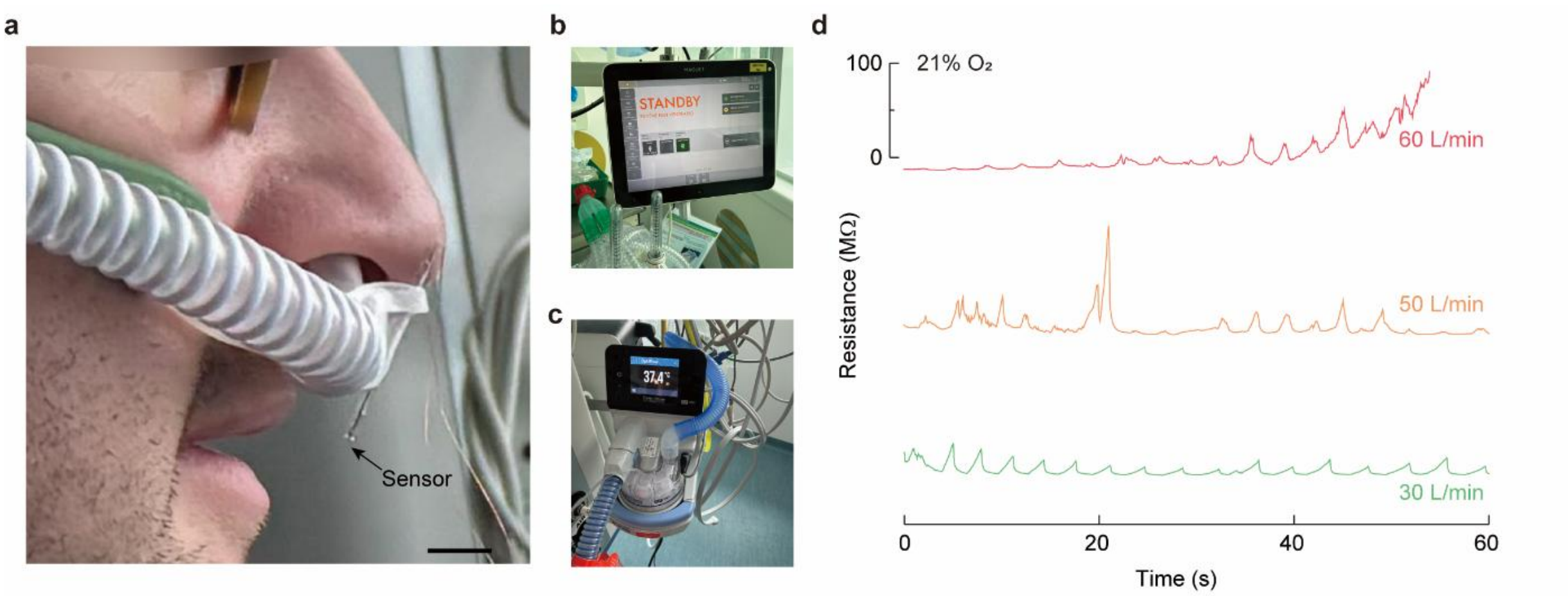

**Supplementary Figures 18 High-flow oxygen therapy breathing measurements and set-up.** (a) Photograph of the single-thread sensor integrated into a high-flow oxygen cannula through a capillary tube. Scale bar: 1 cm. (b) User interface for controlling the gas flow rate and oxygen concentration during the high-flow measurements. (c) Photograph of the humidifier used in the high-flow oxygen delivery system. (d) Representative breathing waveforms recorded under different gas flow rates of 30, 50 and 60 L min$^{-1}$.

Owing to the substantial inspiratory flow delivered by the HFOT system, which typically necessitates mouth breathing, the experimental setup was adapted to ensure accurate signal acquisition. The single-thread MOF sensor was integrated into a fine capillary tube, which served as a structural support to precisely position the sensing element at a fixed distance in front of the participant's mouth (see Figure 4g). This configuration allowed for the continuous monitoring of respiratory cycles superimposed on the constant high-flow baseline. Data acquisition was performed using an RS PRO RSDM3055 Bench Digital Multimeter to record the resistance changes in real-time.

The experimental protocol focused on assessing the sensor's stability across varying oxygen concentrations and flow rates ranging from 30 L/min to 60 L/min. This ensured that the single-thread MOF sensor could consistently distinguish individual respiratory cycles despite the continuous, near-saturated humidity background and high-flow baselines characteristic of the clinical system. Furthermore, the volunteer performed specific tasks including talking and walking while connected to the HFOT system to simulate real-world patient monitoring. These dynamic tests were designed to evaluate the sensor's ability to maintain a clear respiratory signal and a stable baseline during physical movement and vocal turbulence, which are common challenges for wearable sensors in a clinical ward environment. The findings confirmed that the single-thread humidity sensor effectively preserved signal integrity against motion-induced noise and constant high-flow interference.

## Supplementary Note 12. Humidity sensor array test

### Controlled humidity calibration setup

To calibrate the sensor array, a static controlled humidity environment was established using the oversaturated salt solution method within a bespoke sealed chamber (Supplementary Fig. 19). The array was positioned directly above 80 mL of oversaturated salt solution, while a high-precision commercial humidity sensor (SHT85, Sensirion) was placed near the array to provide a reference. The chamber remained hermetically sealed throughout the process to ensure steady-state vapor concentration. The SHT85 sensor was interfaced with an Arduino microcontroller (UNO R3) to stream reference data *via* serial communication to a laptop. Reference humidity values collected during the quasi-steady state were subsequently applied for sensor calibration.

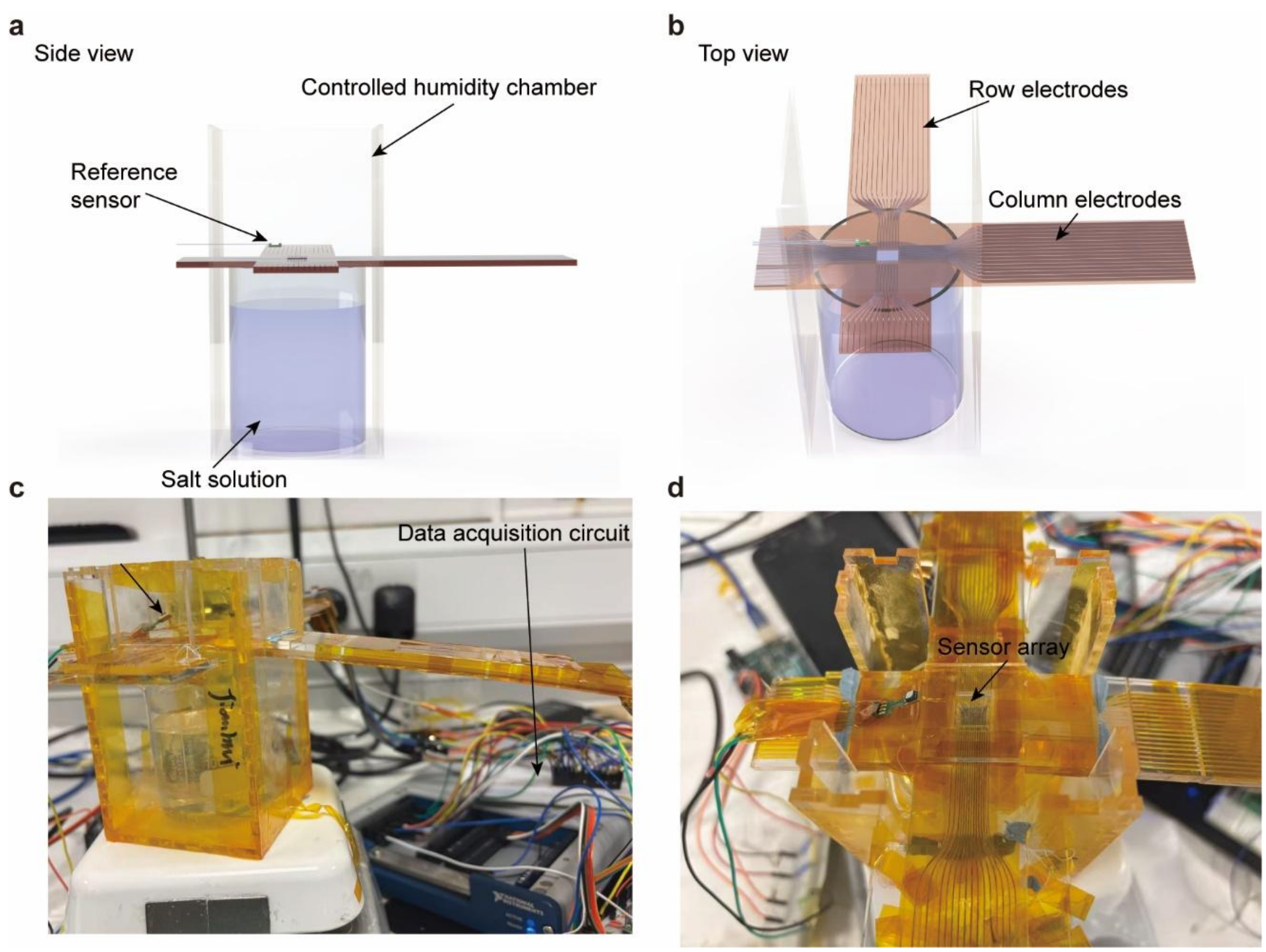

**Supplementary Figures 19 Calibration chamber for controlled-humidity measurements of the sensor array.** (a, b) Schematics of the setup in (a) side view and (b) top view. (c) Side-view photograph of the calibration chamber used for static humidity calibration. (d) Top-view photograph of the chamber without the cap. These figures show the SHT85 reference sensor, oversaturated salt solution, sensor array, row and column electrodes, and DAQ connections.

### Readout electronics and array acquisition architecture

The humidity imaging platform comprises a 16 × 16 sensor array using a row-column addressing architecture (Supplementary Fig. 20). Rows are sequentially enabled *via* a multiplexer (MUX, ADG1606 16:1), while the 16 column signals are read out in parallel through a bank of 16 transimpedance amplifiers (TIAs). The analog front-end is constructed using electrometer-grade operational amplifiers (ADA4530-1), selected for their ultra-low input bias current and minimal leakage, which are critical for the stable readout of high-impedance protonic sensing elements. Each TIA stage incorporates a 10 MΩ feedback resistor to convert the femto-to-nanoampere range sensor currents into measurable voltages. The entire readout electronics system is interfaced with a Data Acquisition (DAQ) card (National Instruments USB-6341), which provides synchronised analog output for excitation, analog input for sampling, and digital logic for row addressing. The control software was implemented in Python within the Anaconda environment to facilitate seamless integration with downstream machine learning frameworks.

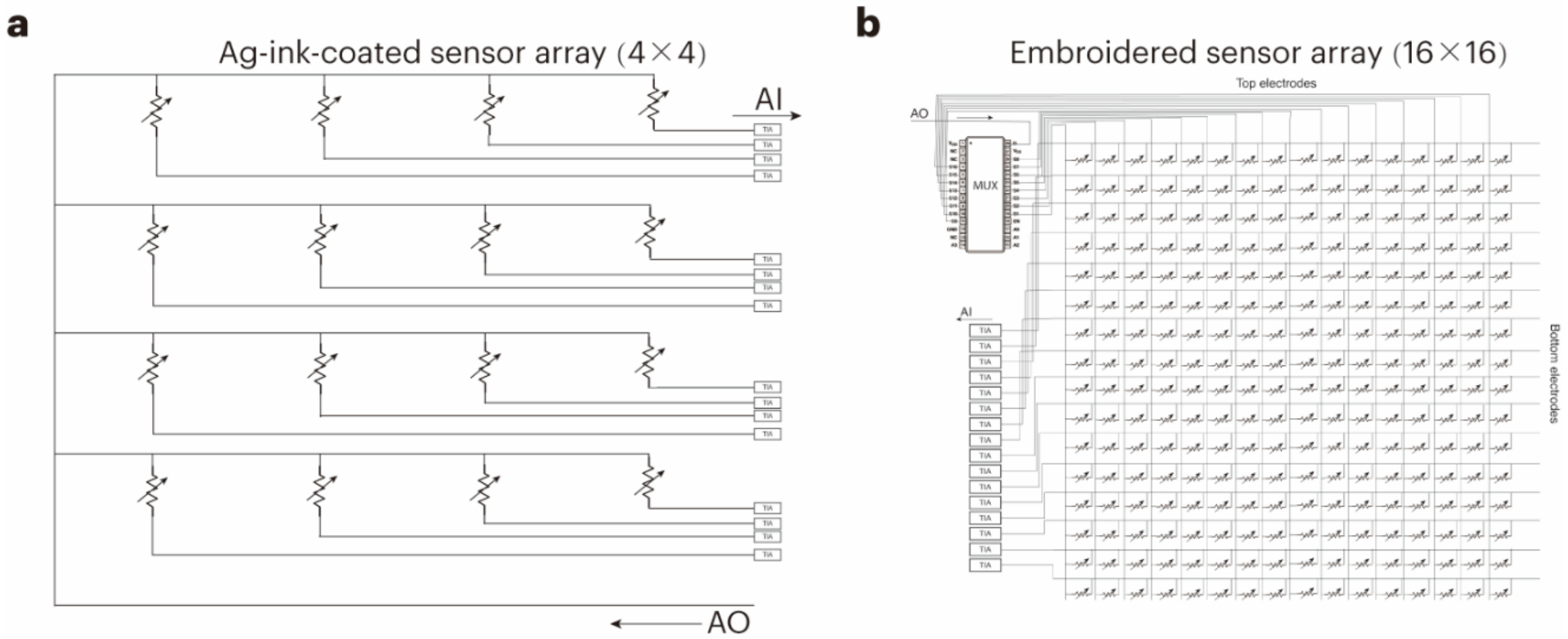

**Supplementary Figures 20 Readout circuit schematics for the textile humidity-sensing arrays.** (a) Circuit schematic of the 4 × 4 Ag-ink-coated sensor array. (b) Circuit schematic of the embroidered high-density sensor array, showing the TIA and MUX. The analog output (AO) and analog input (AI) channels were controlled and acquired by the DAQ card.

During array acquisition, a negative excitation voltage (-1 V) is applied to the active row while the remaining rows remain undriven. The DAQ digital output lines control the multiplexer enable and address pins to scan all 16 rows consecutively. The 16 column outputs are sampled simultaneously by the DAQ's analog input channels, such that a complete frame captures all 256 pixels. In standard full-frame mode, the analog excitation and row-addressing waveforms are pre-generated and hardware-timed to ensure deterministic synchronisation between row gating, excitation, and analog-to-digital conversion. The system operates at a sampling rate of 3000 Hz, with 60 samples acquired per row. To suppress transient artifacts associated with capacitive charging and multiplexer settling times, the first 5 samples following each row switch are discarded.

**Transient waveform-to-humidity regression *via* CalNet**

To convert the raw time-domain sensor responses into relative humidity prediction, we developed a regression network for array calibration, termed CalNet (Figure 5g). Unlike single-thread, single-node measurements, array readout involves row switching through a multiplexer, resulting in pronounced transient responses. These transients evolve rapidly but exhibit reproducible RC-like decay characteristics, allowing humidity-relevant features to be extracted from the transient waveforms using a neural network model.

The input to CalNet was a four-dimensional waveform tensor comprising the batch dimension, the number of temporal samples for each pixel, and the two spatial dimensions corresponding to the 16 rows and 16 columns of the array. This tensor was then rearranged into 256 independent pixel waveforms, each processed by an independent temporal encoder. The encoders were implemented using grouped one-dimensional convolutions with 256 groups. This convolutional design ensured that temporal features from different pixels were not mixed.

The encoder contained four temporal convolutional stages. Each stage consisted of a dilated one-dimensional convolutional layer (Conv1d) block followed by a non-dilated Conv1d block. The convolution kernel size was 3, the per-pixel channel width was 8, and the dilation factors of the four stages were 1, 2, 4 and 8, respectively. Each convolutional block comprised Conv1d, group normalisation, sigmoid linear unit (SiLU) activation and dropout. No intermediate pooling layers were used.

Temporal feature aggregation was performed only once, at the end of the encoder, using statistical pooling along the temporal dimension. For each pixel, the pooled feature vector consisted of the temporal mean, standard deviation, maximum and 95th percentile of the encoded features. These pooled features were then passed to independent two-layer multilayer perceptron (MLP) heads for each pixel to map them to the corresponding humidity value. Each head had a hidden dimension of 16 and used SiLU activation with a dropout of 0.05.

This architecture was designed to reduce the output-collapse problem encountered when fully shared regressors are trained using spatially uniform humidity labels. By retaining independent temporal encoders and output heads for each pixel, CalNet preserves pixel identity and accounts for persistent pixel-to-pixel differences in responsivity, gain and waveform morphology, while still being trained using a common humidity supervision signal.

CalNet was trained using the AdamW optimiser with a learning rate of $2\times10^{-3}$, a batch size of 128 frames and SmoothL1 loss. The model was trained for 80 epochs using automatic mixed precision on a CUDA GPU. To improve performance in the high-humidity regime, the per-pixel loss was weighted by a sigmoid function of the target mean humidity. To reduce the tendency of the model to learn spatially uniform outputs, soft-mask mixing augmentation was used during training.

**Super-resolution reconstruction**

To visualise continuous humidity fields, the 16×16 humidity maps predicted by CalNet were further upsampled using a small two-dimensional super-resolution (SR) network. The SR model first upsampled the low-resolution input by bilinear interpolation and then refined the interpolated map using a residual convolutional module. This module contained eight residual blocks, each with 64 feature channels. Each residual block consisted of two 3×3 convolutional layers, with ReLU activation after the first convolution and a local residual connection. Finally, a 3×3 convolutional layer generated a refinement map, which was added to the bilinearly upsampled input through a global residual connection.

During inference, reflective padding followed by crop-back reconstruction was used to reduce boundary artefacts. The SR output was used only for qualitative visualisation of the humidity field. All quantitative analyses below were performed on the CalNet outputs before super-resolution.

**Model implementation and training environment**

All neural networks were implemented in Python 3.11.15 using PyTorch with CUDA acceleration. Training and inference were performed on a laptop equipped with an NVIDIA GeForce RTX 5080 Laptop GPU (16 GB memory).

**Quantitative analysis and reproducibility of error metrics**

For each acquired frame, CalNet produced a predicted humidity map $\widehat{H}$ . The synchronised reference humidity for that frame was denoted by $H_{ref}$. The Bland–Altman bias heat map quantifies the systematic prediction offset at each pixel. For each pixel $(i, j)$, the bias was calculated as the mean signed difference between the predicted humidity and the synchronised reference humidity across all valid frames $f$:

$$Bias(i,j) = \frac{1}{N}\sum_{f=1}^{N}\left[\widehat{H}_f(i,j) - H_{ref,f}\right] \quad (13)$$

This representation is directly analogous to the bias term in Bland–Altman analysis, but spatially resolved across the sensor array. To quantify the spatial distribution of prediction precision while accounting for the absolute humidity scale, we computed the root-mean-square error for each pixel as

$$RMSE(i,j) = \sqrt{\frac{1}{N}\sum_{f=1}^{N}\left[\widehat{H}_f(i,j) - H_{ref,f}\right]^2} \quad (14)$$

This was converted into a coefficient of variation of the RMSE, denoted CV(RMSE), by normalising the RMSE to the mean reference humidity of the included frames:

$$CV(RMSE)(i,j) = 100 \times \frac{(RMSE)(i,j)}{\overline{H_{ref}}} \quad (15)$$

where $\overline{H_{ref}}$ is the arithmetic mean of the synchronised reference humidities across the same frame set. CV(RMSE) was expressed as a percentage and displayed as a heat map (see Supplementary Fig. 21a). This normalisation allows direct comparison of error magnitude across experiments acquired at different nominal humidity levels.

To quantify how uniformly the array responded at different humidity levels, frames were first grouped according to their synchronised reference humidity. Reference humidities were binned using a fixed bin width ΔH (5% RH), and only bins containing at least a predefined minimum number of frames were retained. For each bin, the spatial mean and spatial standard deviation across the 256 pixels were computed as

$$\mu_f = \frac{1}{256}\sum_{i,j}\widehat{H}_f(i,j) \quad (16)$$

$$\sigma_f = \sqrt{\frac{1}{256}\sum_{i,j}\left(\widehat{H}_f(i,j) - \mu_f\right)^2} \quad (17)$$

The spatial standard deviation $\sigma_f$ quantifies spatial non-uniformity at that instant. For each bin, the signal term was defined as the difference between the bin-mean predicted humidity and the baseline predicted humidity, while the noise term was defined by the mean spatial non-uniformity across frames in that bin. The spatial signal-to-noise ratio (SNR) was then computed as

$$SNR = 20\log_{10}\left(\frac{|\mu_f - \mu_{baseline}|}{\sigma_f}\right) \quad (18)$$

where $\mu_{baseline}$ is the mean predicted humidity of the baseline bin (the lowest RH). The SNR was expressed in dB and plotted as a function of the bin-centre relative humidity (see Supplementary Fig. 21b). This metric captures how strongly the average array response changes with humidity relative to the frame-wise pixel-to-pixel non-uniformity.

**Spatial error characteristics**

The raw CalNet predictions showed only a small systematic offset across the array (Figure 5g and Supplementary Fig. 21). This analysis was performed using 27,109 valid frames, corresponding to 6,939,904 pixel-wise comparisons across the 16×16 array. The Bland–Altman bias remained between 0 and -0.8% RH at the pixel level, indicating that the calibration direction was correct and no large global bias remained. The residual errors were mainly small, spatially structured differences between individual pixels.

These stable spatial offsets likely originate from variations in the electrical junctions where the silver yarns touch the humidity-sensitive fibres. Specifically, differences in contact geometry can change the contact resistance and interfacial capacitance, affecting the gain and transient response of each sensing unit. Some row- or column-dependent non-uniformity might also come from residual variations in the readout circuitry and parasitic wiring. Consistent with this, the CV(RMSE) map remained low (~0.05–0.12) but showed persistent differences in precision across the array.

Spatial SNR analysis showed that at low humidity (near 48% RH), the array response was hard to distinguish from the baseline, with SNR values below 0 dB. However, the SNR increased significantly with humidity, reaching nearly 20 dB at 100% RH. This confirms that the imaging system becomes much more reliable as humidity contrast increases. Despite this, the effective operating range of the integrated array was narrower than that of a single-fiber sensor. This suggests that current limitations arise from array integration and multi-channel readout variability, rather than the intrinsic sensitivity of the MOF nanostrip material itself.

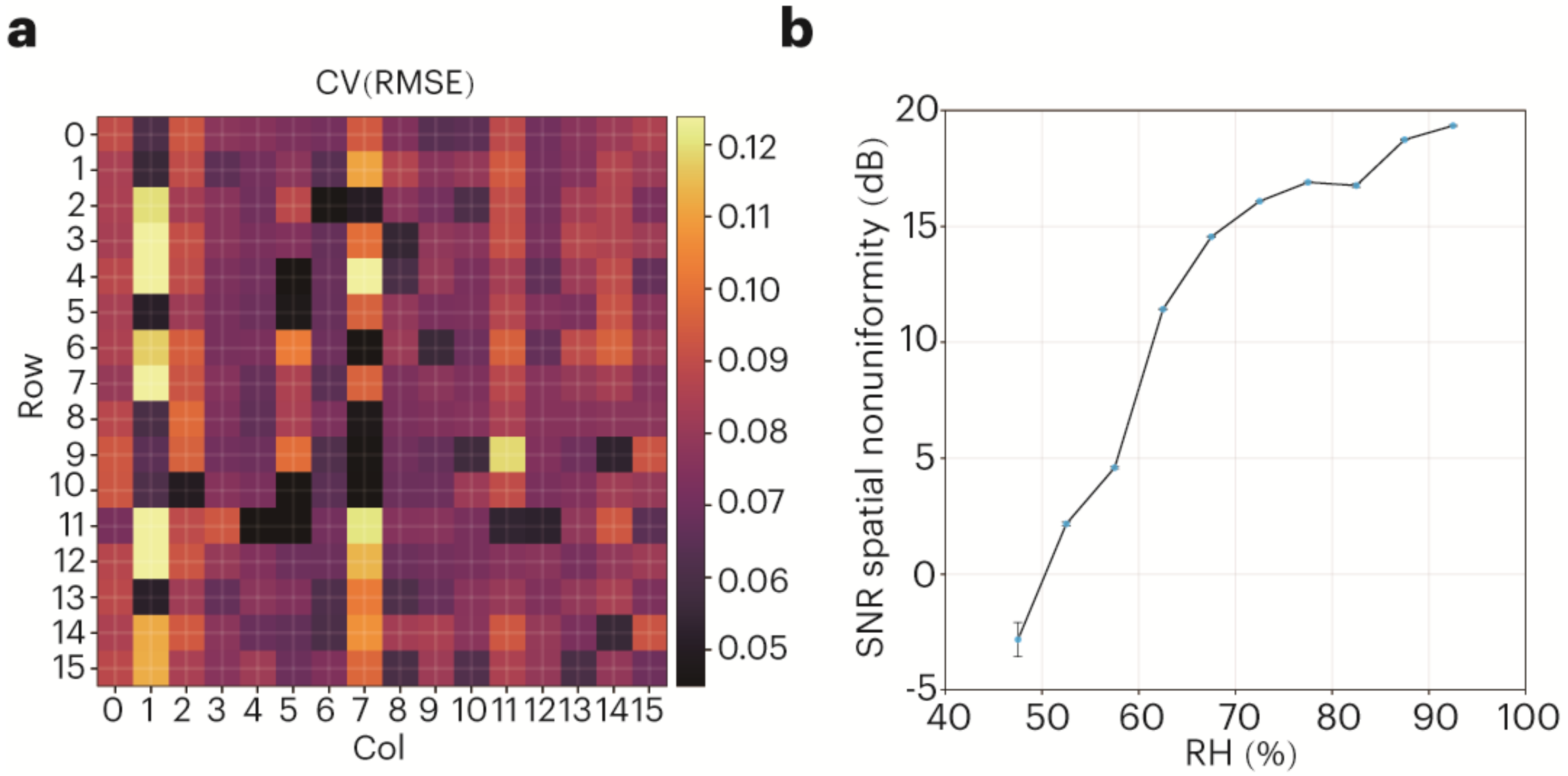

**Supplementary Figures 21 Calibration and signal-quality analysis of the 16 × 16 humidity-sensing array.** (a) Heatmap of the coefficient of variation of the root mean squared error across the array. (b) SNR plotted as a function of the bin-centre relative humidity.

## Supplementary Note 13. *Ab Initio* molecular dynamics simulations

*Ab initio* molecular dynamics (AIMD) simulations based on density functional theory (DFT) were performed to investigate the behaviour of water molecules and protons both within the MOF crystal and on its surface [31]. All simulations were conducted using the CP2K/Quickstep package, employing the Perdew–Burke–Ernzerhof (PBE) exchange–correlation functional with Goedecker–Teter–Hutter (GTH) pseudopotentials and a double-ζ valence polarised (DZVP) basis set. A hybrid Gaussian and plane wave (GPW) scheme was used to compute the electronic ground-state structure, with a plane-wave cutoff of 300 Ry for the electron density. Empirical dispersion corrections were included via the D3 method. Two simulation models were constructed to represent different humidity conditions: the low-humidity model included five water molecules and one proton, while the high-humidity model contained an additional 14 water molecules. Prior to AIMD simulations, gas-phase geometry optimisations were performed at the DFT-PBE level. The self-consistent field (SCF) energy convergence criterion was set to $10^{-6}$. AIMD simulations were carried out in the canonical (NVT) ensemble for 20 picoseconds (ps) using a 1 femtosecond (fs) time step. Temperature was maintained at 298 K using the Nosé–Hoover thermostat.

## Captions for supplementary movies

Supplementary movie 1: Robustness test. The tests were performed to evaluate the stability of the sensor, including high-speed jet impact, cyclic bending test, washing machine test, and alcohol immersion test.

Supplementary movie 2: Nasal breathing monitoring with single-thread breathing sensor embedded in a nasal cannula.

Supplementary movie 3: Oral and nasal breathing monitoring using a single-thread breathing sensor.